\documentclass[11pt,American,A4paper,twoside]{article}
\newif\ifdraft \global\drafttrue
\def\production{\global\draftfalse}
\production
\usepackage[T1]{fontenc}
\usepackage[latin1]{inputenc}
\usepackage{a4wide}
\usepackage{times}
\usepackage{graphicx}
\usepackage{theorem}
\usepackage[colorlinks=true,hyperindex=true]{hyperref}
\usepackage{color}
\usepackage{fancyhdr}
\usepackage{latexsym}
\usepackage{amsmath}
\usepackage{bbm}
\usepackage{amssymb}
\usepackage{amsfonts}
\usepackage{enumerate}
\usepackage{cancel}
\usepackage{stmaryrd}
\ifdraft
\usepackage{showkeys}
\fi
\newcounter{smallarabics}

\newcounter{smallroman}

\newcommand{\ben}{\begin{enumerate}[{\rm (1)}]}
\newcommand{\een}{\end{enumerate}}
\newcounter{enumisaved}
\newtheorem{theoreme}{Theorem }[section]
\newtheorem{proposition}[theoreme]{Proposition}
\newtheorem{lemma}[theoreme]{Lemma}
\newtheorem{definition}[theoreme]{Definition}
\newtheorem{corollary}[theoreme]{Corollary}

\def\bep{\begin{proposition}}
\def\eep{\end{proposition}}

\def\bel{\begin{lemma}}
\def\eel{\end{lemma}}
\def\bet{\begin{theoreme}}
\def\eet{\end{theoreme}}
\def\bed{\begin{definition}}
\def\eed{\end{definition}}
\def\bec{\begin{corollary}}
\def\eec{\end{corollary}}

\def\rr{{\mathbb R}}

\def\nn{{\mathbb N}}

\def\textsl{{}}
\let\oldi\i
\newcommand{\myi}{\oldi}

\def\c0inf{C_0^\infty}

\def\proof{\noindent {\bf Proof.}\ \ }

\def\cH{{\cal  H}}

\def\cA{{\cal A}}
\def\PP{\mathbb{P}}
\def\EE{\mathbb{E}}
\def\QQ{\mathbb{Q}}

\def\cB{{\cal B}}

\def\i{{\rm i}}

\newcommand{\beq}{\begin{equation}}
\newcommand{\eeq}{\end{equation}}
\newcommand{\bear}[1]{\begin{array}{#1}}
\newcommand{\ear}{\end{array}}

\def\sp{{\hat e}}

\newcommand{\e}{\mathrm{e}}
\renewcommand{\i}{\mathrm{i}}

\renewcommand{\d}{\mathrm{d}}

\newcommand{\ep}{\mathrm{ep}}

\def\qed{$\Box$\medskip}

\def\cP{{\cal P}}
\def\cJ{{\cal J}}
\def\cC{{\cal C}}
\def\cT{{\cal T}}
\def\cF{{\cal F}}
\def\cI{{\cal I}}

\def\cA{{\cal A}}

\def\bar{\overline}
\def\ubar{\underline}

\def\12{\frac{1}{2}}

\def\supp{{\rm supp}}

\def\e{{\rm e}}
\def\cD{{\cal D}}

\def\d{{\rm d}}

\def\one{{\mathbbm 1}}

\def\cH{{\cal H}}

\def\sp{{\rm sp}}

\def\cX{{\cal X}}
\def\cU{{\cal U}}

\def\eq{{\rm eq}}

\def\tr{{\rm tr}}

\def\P{\mathbb P}
\def\cPi{{\mathcal{P}_{\phi}}}
\def\cO{{\cal O}}
\newcommand{\ds}{\displaystyle}

\def\rhoi{{\rho_{\rm inv}}}
\def\wP{{\widehat{\mathbb P}}}

\def\bomega{\boldsymbol{\omega}}
\def\bnu{\boldsymbol{\nu}}
\def\bxi{\boldsymbol{\xi}}
\def\Beta{\boldsymbol{\eta}}
\begin{document}
\def\today{}
\title{On entropy production of repeated quantum measurements I.\\
General theory }
\author{ T. Benoist$^{1}$, V. Jak\v{s}i\'c$^{2}$, Y. Pautrat$^{3}$,  C-A. Pillet$^{4}$
\\ \\ \\
$^1$ CNRS, Laboratoire de Physique Th\'eorique, IRSAMC,\\
Universit\'e de Toulouse, UPS, F-31062 Toulouse, France
\\ \\
$^2$Department of Mathematics and Statistics, 
McGill University, \\
805 Sherbrooke Street West, 
Montreal,  QC,  H3A 2K6, Canada
\\ \\
$^3$Laboratoire de Math\'ematiques d'Orsay, Univ. Paris-Sud, CNRS,\\ Universit\'e
Paris-Saclay, 91405 Orsay, France
%$^3$Laboratoire de Math\'ematiques, Universit\'e Paris-Sud, 91405 Orsay Cedex, France
\\ \\
$^4$Aix Marseille Univ, Univ Toulon, CNRS, CPT, Marseille, France
}

\maketitle

\bigskip
\centerline{\large \bf Dedicated to the memory of Rudolf Haag}

\bigskip
\bigskip
\bigskip

{\small
{\bf Abstract.} We study entropy production (EP) in processes involving repeated
quantum measurements of finite quantum systems. 
Adopting a dynamical system approach, we develop
a thermodynamic formalism for the EP and study fine aspects of irreversibility 
related to the hypothesis testing of the arrow of time.
Under a suitable chaoticity assumption, we  establish a Large Deviation
Principle and a Fluctuation Theorem for the EP.
}

\thispagestyle{empty}
%%%%%%%%%%%%%%%%%%%%%%%%%%%%%%%%%%%%%%%%%%%%%%
%%%%%%%%%%%%%%%%%%%%%%%%%%%%%%%%%%%%%%%%%%%%%%

\tableofcontents

\section{Introduction} 
\label{sec-intro}
\subsection{Historical perspective}

In 1927, invoking the "dephasing effect" of the interaction of a system with a
measurement apparatus, Heisenberg~\cite{He} introduced the ``reduction of the
wave function'' into quantum theory as the  proper way to assign a wave function
to a quantum system after a successful measurement. In the same year,
Eddington~\cite{Ed} coined the term "arrow of time" in his discussion of the
various aspects of irreversibility in physical systems. Building on Heisenberg's
work,  in his  1932 monograph~\cite{VN}, von Neumann developed the first
mathematical theory of quantum measurements. In this theory, the wave function
reduction leads to an intrinsic irreversibility of the measurement process that
has no classical analog and is sometimes called the {\em quantum arrow of time}.
Although a consensus has been reached around the so-called
orthodox\footnote{sometimes facetiously  termed  the  "shut up and calculate"
approach; see~\cite{MND}.} approach to quantum measurements~\cite{Wi}, after
nearly a century of research, their fundamental status within quantum mechanics
and their problematic  relationship with ``the observer'' 
are far from understood and remain much  debated;
see~\cite{HMPZ,ST,Ze,BFFS,BFS}.

Regarding the quantum arrow of time, Bohm~\cite[Section~22.12]{Bo} points out that {\em 
"[quantum] irreversibility greatly resembles that which appears in thermodynamic
processes,"} while Landau and Lifshitz~\cite[Section~I.8]{LL} go further and
discuss the possibility  that the second law of thermodynamics and  the
thermodynamical arrow of time are macroscopic expressions of the quantum arrow of
time. In 1964, Aharonov, Bergmann and Lebowitz~\cite{ABL}  critically
examined the nature of the quantum arrow of time. They showed that conditioning
on both the initial and final quantum states, one could construct a time symmetric
statistical ensemble of quantum measurements, lifting the problem of  time
irreversibility implied by the projection postulate to a question of appropriate
choice of a statistical ensemble. The construction of this "two-state vector
formalism"~\cite{AV}  and weak measurements~\cite{BG,Ca,Da,WM} has led to the
definition of weak values by Aharonov, Albert and Vaidman~\cite{AAV} which in
turn played an important role in recent developments in quantum
cosmology~\cite{ST}.

Independently, and on a more pragmatic ground, new ideas have emerged in the
study of nonequilibrium processes~\cite{Ru2}. Structured by the concepts of
nonequilibrium steady state and entropy production, they have triggered 
intense activity in both theoretical and experimental physics. In the resulting
theoretical framework, classical fluctuation relations~\cite{ECM,GC1,GC2,Jar,Cr1} 
hint at new links between the thermodynamic formalism of dynamical
systems~\cite{Ru1} and statistical mechanics. In this formalism,  the time arrow is intimately 
linked with information theoretic concepts and  its
emergence can be precisely quantified. Fluctuation relations have been extended
to quantum dynamics~\cite{Ku2,Ta,EHM,CHT,DL,JOPP}, allowing for the
study of the  arrow of time in open quantum systems~\cite{JOPS,BSS}.

While repeated quantum measurement processes have recently received much
focus~\cite{SVW,GPP,YY,BB,BBB,BFFS},  their connections
with the above mentioned advances of nonequilibrium statistical mechanics
have not been fully explored. This work is our first
step in this direction of research. It aims at a better understanding of the
large time asymptotics of the statistics of the fluctuations of entropy 
production in repeated quantum measurements. We shall in particular derive the large deviation principle for these fluctuations and 
prove the so-called fluctuation theorem.
We shall also quantify the emergence of the arrow of time by  linking 
hypothesis testing exponents distinguishing past and future  to the large deviation principle for fluctuations of entropy production. 

Finally, we stress that even though the systems of interest in this work are of a 
genuine quantum nature, the resulting sequences of measurement outcomes are 
described by classical dynamical systems. However, these dynamical systems do not 
generally satisfy the "chaotic hypothesis" of Gallavotti-Cohen\footnote{The relevant invariant measures are non-Gibbsian.}. From a  mathematical  perspective, the main results
of this work are extensions of the thermodynamic formalism to this new class
of dynamical systems.

\subsection{Setting}

We shall focus on measurements described by a (quantum) {\sl instrument} $\cJ$
on a finite dimensional Hilbert space $\cH$. 
We briefly recall the corresponding setup, referring the reader
to~\cite[Section~5]{Ho} for additional information and references to the
literature, and to~\cite[Section~4]{Da} for a discussion of pioneering works on
the subject.

An instrument in the Heisenberg picture is a finite family 
$\cJ=\{\Phi_a\}_{a\in\cA}$
of completely positive maps $\Phi_a:\cB(\cH)\to\cB(\cH)$\footnote{$\cB(\cH)$
denotes the algebra of all linear maps $X:\cH\to\cH$.} such that
$\Phi:=\sum_{a\in\cA}\Phi_a$ satisfies $\Phi[\one]=\one$. The finite alphabet
$\cA$  
 describes the possible outcomes of a single measurement.
We denote by $\cJ^\ast=\{\Phi_a^\ast\}_{a\in\cA}$ the dual instrument in the
Schr\"odinger picture, where $\Phi_a^\ast:\cB(\cH)\to\cB(\cH)$ is defined by
$\tr(\Phi_a^\ast[X]Y)=\tr(X\Phi_a[Y])$.

A pair $(\cJ,\rho)$, where $\cJ$ is an instrument and $\rho$ a density matrix on $\cH$,
defines a repeated measurement process in the following way.
At time $t=1$, when the system is in the state $\rho$, a 
measurement is performed. The outcome $a_1\in\cA$ is observed with probability 
$\tr(\Phi_{a_1}^\ast[\rho])$, and after the measurement the system is in the state 
\[
\rho_{a_1}=\frac{\Phi_{a_1}^\ast[\rho]}{\tr(\Phi_{a_1}^\ast[\rho])}.
\]
A further measurement at time $t=2$ gives the outcome $a_2$ with probability
\[
\tr(\Phi_{a_2}^\ast[\rho_{a_1}])
=\frac{\tr((\Phi_{a_2}^\ast\circ\Phi_{a_1}^\ast)[\rho])}{\tr(\Phi^\ast_{a_1}[\rho])},
\]
and the joint probability for the occurence of the sequence of outcomes $(a_1,a_2)$ is
\[
\tr ((\Phi_{a_2}^\ast\circ\Phi_{a_1}^\ast)[\rho])
=\tr(\rho\,(\Phi_{a_1}\circ\Phi_{a_2})[\one]).
\]
Continuing inductively, we deduce that the distribution of the sequence 
$(a_1,\ldots,a_T)\in\cA^T$ of outcomes of $T$ successive measurements is given by 
$$
\PP_T(a_1,\ldots,a_T)
=\tr\left(\rho\,(\Phi_{a_1}\circ\cdots\circ\Phi_{a_T})[\one]\right).
$$
One easily verifies that, due to the relation $\Phi[\one]=\one$, $\PP_T$ is 
indeed a probability measure on $\cA^T$. 

\begin{quote} {\bf Example 1. von Neumann measurements.} Suppose that ${\cal A}\subset \rr$. Let $\{P_a\}_{a\in\cA}$ be a 
family of orthogonal projections satisfying $\sum_{a\in\cA} P_a=\one$, and let the
unitary $U:\cH\to\cH$ be the propagator of the system over a unit time 
interval. The instrument defined by $\Phi_a[X]=V_a XV_a^\ast$, where  $V_a=U^\ast P_a$, 
describes the projective von Neumann measurement of the observable 
$A=\sum_{a\in\cA} aP_a$. More precisely, if the system is in the state $\rho$ at
time $t$, then a measurement of $A$ at time $t+1$ yields $a$ with the probability
$\tr(\Phi_a^\ast[\rho])$. 

\end{quote}
\begin{quote} {\bf Example 2. Ancila measurements.} 
Let $\cH_p$ be a finite dimensional Hilbert space describing a "probe" allowed to 
interact with our system. The initial state of the probe is described by the density 
matrix $\rho_p$. The Hilbert space and the initial state of the joint system are  
$\cH\otimes\cH_p$ and $\rho\otimes\rho_p$. Let  $U:\cH\otimes\cH_p\to\cH\otimes\cH_p$ be the unitary propagator of the joint system
over a unit time interval. Let $\{P_a\}_{a\in\cA}$ be a family of orthogonal projections 
on $\cH_p$ such that $\sum_{a\in\cA}P_a=\one$ and define
\beq
\Phi_a^\ast[\rho]=\tr_{\cH_p}\left((\one\otimes P_a)U(\rho\otimes\rho_p)U^\ast\right),
\label{flash}
\eeq
where $\tr_{\cH_p}$ stands for the partial trace over $\cH_p$. The maps $\Phi_a^\ast$ 
extend in an obvious way to linear maps $\Phi_a^\ast:\cB(\cH)\to\cB(\cH)$, and the 
family $\{\Phi_a^\ast\}_{a\in\cA}$ is an instrument on $\cH$ in the Schr\"odinger 
picture.\footnote{The same instrument in the Heisenberg picture is described by 
$\Phi_a[X]=\tr_{\cH_p}\left(U^\ast(X\otimes P_a)U(\one\otimes\rho_p)\right)$.}
Moreover, {\em any} such instrument arises in this way: given 
$\{\Phi_a^\ast\}_{a\in\cA}$, one can find $\cH_p,\rho_p$, $U$, and $\{P_a\}_{a\in\cA}$
so that~\eqref{flash} holds for all density matrices $\rho$ on $\cH$.\footnote{In our setting, this result is 
an immediate consequence of Stinespring's dilation theorem. For generalizations, see 
~\cite[Section~5]{Ho}.}
\end{quote}
\begin{quote}{\bf Example 3. Perfect instruments.} An instrument 
$\cJ=\{\Phi_a\}_{a\in\cA}$ is called perfect if, for all $a\in\cA$, 
$\Phi_a[X]=V_a X V_a^\ast$ for some $V_a\in\cB(\cH)$. The instruments 
associated to von Neumann measurements are perfect. The instrument of an 
ancila measurement is perfect if $\dim P_a=1$ for all $a\in\cA$ and $\rho_p$ is a pure 
state, i.e., $\rho_p=|\psi_p\rangle\langle \psi_p|$ for some unit vector 
$\psi_p\in \cH_p$.
\end{quote}

\begin{quote}{\bf Example 4. Unraveling of a quantum channel.} Let 
$\Phi:\cB(\cH)\to\cB(\cH)$ be a completely positive unital map -- a quantum channel.
Any such $\Phi$ has a (non-unique) Kraus representation
$$
\Phi[X]=\sum_{a\in\cA} V_a XV_a^\ast,
$$
where $\cA$ is a finite set and the $V_a\in\cB(\cH)$ are such that 
$\sum_{a\in\cA} V_a V_a^\ast=\one$ (see, e.g., \cite[Theorem~2.2]{Pe}).
The maps $\Phi_a[X]=V_a XV_a^\ast$ define 
a perfect instrument. The process $(\cJ,\rho)$ induced by the Kraus family 
$\cJ=\{\Phi_a\}_{a\in\cA}$ and a state $\rho$ is a so-called unraveling of $\Phi$.
\end{quote}

The property $\Phi[\one]=\one$ ensures that the family $\{\P_T\}_{T\geq1}$ uniquely 
extends to a probability measure $\P$ on $(\Omega,\cF)$, where $\Omega =\cA^\nn$ and 
$\cF$ is the $\sigma$-algebra on $\Omega$ generated by the cylinder sets. $\EE[\,\cdot\,]$  denotes the expectation w.r.t.\;this measure.
We equip 
$\Omega$ with the usual product topology. $\Omega$ is metrizable and a convenient 
metric for our purposes is $d(\omega, \omega^\prime)=\lambda^{k(\omega, \omega^\prime)}$, 
where $\lambda \in ]0,1[$ is fixed and $k(\omega, \omega^\prime)=\inf\{j\,|\, \omega_j\not=\omega_j^\prime\}$;  see 
\cite[Section~7.2]{Ru1}.
$(\Omega, d)$ is a compact metric space and its Borel $\sigma$-field coincides 
with $\cF$. For any integers $1\le i\le j$ we set $\llbracket i,j\rrbracket=[i,j]\cap\nn$
and denote $\Omega_T=\cA^{\llbracket1,T\rrbracket}$. The left shift 
$$
\begin{array}{rccc}
\phi:&\Omega&\to&\Omega\\
&(\omega_1,\omega_2,\ldots)&\mapsto&(\omega_2,\omega_3,\ldots),
\end{array}
$$ 
is a continuous surjection. If the initial state $\rho$ satisfies 
$\Phi^\ast[\rho]=\rho$,\footnote{Since $\Phi[\one]=\one$, such a density 
matrix $\rho$ always exists.} then  $\P$ is $\phi$-invariant (i.e., 
$\P(\phi^{-1}(A))=\P(A)$ for all $A\in\cF$) and  the process $(\cJ,\rho)$ defines a 
dynamical system $(\Omega,\P,\phi)$. This observation  leads to our first 
assumption:
\begin{quote}\label{A-def}
{\bf Assumption (A)} The initial state  satisfies $\Phi^\ast[\rho]=\rho$
and $\rho >0$. 
\end{quote}
\newcommand{\AssumptionA}{{\hyperref[A-def]{{\rm (A)}}}}%
In what follows we shall always assume that~\AssumptionA{} holds. Some of our 
results hold under a weaker form of Assumption~\AssumptionA; 
see Remark~2 after Theorem~\ref{stein}.

The  basic ergodic properties of dynamical system $(\Omega, \P, \phi)$  can be characterized in terms of $\Phi$ 
as follows. Note that the spectral radius of $\Phi$ is $1$. 

\bet\label{thm-km}
\begin{enumerate}[{\rm (1)}] 
\item If $1$ is a simple eigenvalue of $\Phi$, then $(\Omega, \P, \phi)$ is ergodic.
\item If $1$ is a simple eigenvalue of $\Phi$ and $\Phi$ has no other eigenvalues on 
the unit circle $|z|=1$, then $(\Omega, \P, \phi)$  is a $K$-system, and in particular it is mixing. Moreover, 
 for any two H\"older continuous functions $f, g : \Omega\rightarrow \rr$ there exists a constant $\gamma >0$ such that
\[\EE(f g\circ \phi^n)-\EE(f)\EE(g)=O(\e^{-\gamma n}).
\]
\end{enumerate}
\eet
{\bf Remark 1.} These results can be traced back to~\cite{FNW}; see Remark~1 in 
Section~\ref{Sec2Rem}.
Related results can be found in~\cite{KM1,KM2, MP}. For a pedagogical 
exposition of the proofs and additional information we refer the reader to \cite{BJPP3}.

{\bf Remark 2.} $1$ is a simple eigenvalue of $\Phi$ whenever  $\Phi$ is irreducible, i.e., the relation
$\Phi[P]\le\lambda P$ for some orthogonal
projection $P\in\cB(\cH)$ and some $\lambda>0$ implies $P\in\{0,\one\}$; see~\cite[Lemma~4.1]{EHK}. 

{\bf Remark 3.} Note that the condition $\Phi^\ast[\rho]=\rho$ is equivalent to
$$
\sum_{\omega_1,\ldots,\omega_S\in\cA}
\PP_{S+T}(\omega_1,\ldots,\omega_S,\omega_{S+1},\ldots,\omega_{S+T})
=\PP_T(\omega_{S+1},\ldots,\omega_{S+T}),
$$
which is usually seen as an effect of decoherence. Under the Assumptions
of Theorem~\ref{thm-km}~(2) the state $\rho$ satisfying Assumption~\AssumptionA{} 
is unique. Moreover, for any density matrix $\rho_0$, one has
$$
\lim_{S\to\infty}\sum_{\omega_1,\ldots,\omega_S\in\cA}
\tr(\rho_0\Phi_{\omega_1}\circ\cdots\circ\Phi_{\omega_S}\circ
\Phi_{\omega_{S+1}}\circ\cdots\circ\Phi_{\omega_{S+T}})=\PP_T(\omega_{S+1},\ldots,\omega_{S+T}),
$$
i.e., the dynamical system $(\Omega, \P, \phi)$ describes the process
$(\cJ,\rho_0)$ in the asymptotic regime where a long sequence of initial
measurements is disregarded.

\medskip
We proceed to describe the entropic aspects of the dynamical system $(\Omega,\PP,\phi)$ 
generated by the repeated measurement process $(\cJ,\rho)$ that will be our main concern. 
Let $\theta:\cA\to\cA$ be an involution. For each $T\ge1$ we define an involution on 
$\Omega_T$ by
$$
\Theta_T(\omega_1,\ldots,\omega_T)=(\theta(\omega_T),\ldots,\theta(\omega_1)).
$$
A process $(\widehat{\cJ},\widehat\rho\,)$ is called an {\em outcome reversal} (abbreviated OR) 
of the process $(\cJ,\rho)$ whenever the instrument
$\widehat{\cJ}=\{\widehat{\Phi}_a\}_{a\in\cA}$ and the density matrix $\widehat\rho$ 
acting on the same Hilbert space $\cH$ satisfy 
$\widehat{\Phi}^\ast[\widehat{\rho}]=\widehat{\rho}>0$, and the induced probability 
measures 
$$
\widehat{\PP}_T(\omega_1,\ldots,\omega_T)
=\tr\left(\widehat\rho\,(\widehat{\Phi}_{\omega_1}\circ\cdots
\circ\widehat{\Phi}_{\omega_T})[\one]\right)
$$
satisfy
\beq
\widehat{\PP}_T=\mathbb{\PP}_T\circ\Theta_T
\label{holds}
\eeq
for all $T\geq 1$. Such a process always exists and  a canonical choice is 
\beq
\widehat{\Phi}_a(X)=\rho^{-\tfrac12}\Phi_{\theta(a)}^\ast\left[\rho^{\tfrac12}
X\rho^{\tfrac12}\right]\rho^{-\tfrac12}, 
\qquad \widehat \rho=\rho;
\label{crooks}
\eeq
see~\cite{Cr2}.
Indeed, one easily checks that $\{\widehat{\Phi}_a\}_{a\in\cA}$ is an instrument such 
that $\widehat{\Phi}^\ast[\widehat{\rho}\,]=\widehat{\rho}$ and that~\eqref{holds} holds.
Needless to say, the OR process $(\widehat{\cJ},\widehat{\rho}\,)$ need not be unique; 
see~\cite{BJPP2} for a discussion of this point. However, note that the family 
$\{\wP_T\}_{T\ge1}$ defined by~\eqref{holds} induces a unique $\phi$-invariant 
probability  measure $\wP$ on $\Omega$: the OR dynamical system $(\Omega,\wP,\phi)$ is completely 
determined by $(\Omega,\P,\phi)$ and the involution $\theta$.

In the present setting our study of the quantum arrow of time concerns the 
{\em distinguishability} between $(\cJ,\rho)$ and its OR 
$(\widehat{\cJ},\widehat{\rho}\,)$ quantified by the entropic distinguishability of the 
respective probability measures $\PP$ and $\wP$. This entropic distinguishability 
is intimately linked with entropy production and hypothesis testing of the same pairs and 
we shall examine it on two levels:
\begin{description}
\item[Level I:] Asymptotics of relative entropies and mean entropy production rate, 
Stein error exponent.
\item[Level II:] Asymptotics of R\'enyi's relative entropies and fluctuations of entropy 
production, large deviation principle and fluctuation theorem, Chernoff and Hoeffding 
error exponents.
\end{description}

\subsection{Two remarks}
\label{Sec2Rem}

{\bf Remark 1.} The dynamical systems $(\Omega, \P, \phi)$ studied in this paper constitute a 
special class of $C^\ast$-finitely correlated states introduced in the seminal paper~\cite{FNW}. We recall the well-known 
construction.  
Let ${\frak C}_1$, ${\frak C}_2$ be two finite-dimensional $C^\ast$-algebras, $\rho$ a state on ${\frak C}_2$, and 
${\cal E}: {\frak C}_1\otimes {\frak C}_2\rightarrow {\frak C}_2$ a completely positive unital map such that for all $B\in {\frak C}_2$, 
\[
\rho\left({\cal E}(\one_{{\frak C}_1}\otimes B)\right)=\rho(B).
\]
For each $A\in {\frak C}_1$ one defines a map ${\cal E}_A: {\frak C}_2\rightarrow {\frak C}_2$ by setting 
${\cal E}_A(B)={\cal E}(A\otimes B)$. The map 
\[
\gamma_n(A_1\otimes \cdots \otimes A_n)=\rho({\cal E}_{A_1}\cdots {\cal E}_{A_n}(\one_{{\frak C_2}}))
\]
uniquely extends to a state on the tensor product $\bigotimes_{i=1}^n {\frak C}_1^{(i)}$, where ${\frak C}_1^{(i)}$ is a copy of 
${\cal C}_1$. Finally, the family of states $\gamma_n$, $n\in  {\mathbb N}$, uniquely extends to a state $\gamma$ on the  $C^\ast$-algebra 
$\bigotimes_{i\in {\mathbb N}}{\frak C}_1^{(i)}$. The state $\gamma$ is the $C^\ast$-finitely correlated state associated to 
$({\frak C}_1, {\frak C}_2, {\cal E}, \rho)$. 
The pairs $(\Omega, \P)$ that arise in repeated quantum 
measurements of finite quantum systems  correspond precisely to $C^\ast$-finitely correlated states with commutative ${\frak  C}_1$ and 
${\frak C}_2=\cB(\cH)$ for some finite dimensional Hilbert space $\cH$. This connection  will  play an important role   in 
the continuation of this work \cite{BJPP2}. 

In this context we also mention a pioneering  work of Lindblad \cite{Li} 
who studied the entropy of finitely correlated states generated by a non-Markovian adapted sequence of instruments. Since
the main focus of  this paper is  the entropy production, the Lindblad work is only indirectly  related to ours, and we will comment further on it 
in \cite{BJPP2}. 

{\bf Remark 2.}  
It is important to emphasize that the object of our study is the classical 
dynamical system $(\Omega,\PP,\phi)$ and that the thermodynamic formalism of 
entropic fluctuations we will develop here is {\em classical} in nature. The 
quantum origin of the dynamical system $(\Omega,\PP,\phi)$ manifests itself in
the interpretation of our results and in the properties of the measure $\PP$. The 
latter differ significantly from the ones usually assumed in the Gibbsian
approach to the thermodynamic formalism. In particular, the 
Gibbsian theory of entropic fluctuations pioneered in~\cite{GC1,GC2} and further 
developed in~\cite{JPR,MV} cannot be applied to $(\Omega,\PP,\phi)$, and a novel 
approach is needed. We will comment further on this point in 
Section~\ref{sec-differ}. Here we mention only that  the  main technical tool of 
our work is the subadditive ergodic theory of dynamical systems  
developed in unrelated studies of the multifractal analysis of a certain class of 
self-similar sets; see~\cite{BaL,BV,CFH,CZC,Fa,FS,Fe1,Fe2,Fe3,FL,FK,IY,KW}. This 
tool sheds an unexpected light on the statistics of repeated quantum 
measurements. 

\subsection{Organization}
The paper is organized as follows. Sections~\ref{sec-entropies}--\ref{sec-Stein} deal with Level~I: Asymptotics of relative entropies and mean entropy production rate, 
Stein error exponent.
In Section~\ref{sec-entropies}, we fix our notation regarding  various kinds of 
entropies that will appear in the paper. In Section~\ref{sec-ep} we state our results 
concerning the entropy production rate of the process $(\cJ,\rho)$. Stein's error 
exponents are  discussed in Section~\ref{sec-Stein}. Sections~\ref{sec-entropy2}--\ref{sec-hyp} deal with  Level~II: Asymptotics of R\'enyi's relative entropies and fluctuations of entropy 
production, large deviation principle and fluctuation theorem, Chernoff and Hoeffding 
error exponents.   Additional notational conventions and 
properties of entropies are discussed in Section~\ref{sec-entropy2}.
Section~\ref{sec-thermo} is devoted to R\'enyi's relative entropy and its thermodynamic 
formalism. Fluctuations of the entropy production rate, including Large Deviation 
Principles as well as local and global Fluctuation Theorems are stated in 
Section~\ref{SEC-LDP}. In Section~\ref{sec-hyp} we discuss
hypothesis testing and, in particular, the Chernoff and Hoeffding error
exponents. 
The proofs are
collected in Sections~\ref{sec-ep-proofs} and~\ref{sec-re-proofs}.

For reasons of space, the discussion of concrete models of repeated quantum measurements to which 
our results apply will be presented in the continuation of this work \cite{BJPP1}.

\bigskip\noindent
{\bf Acknowledgments.} We are grateful to Martin Fraas, J\"urg Fr\"ohlich and 
Daniel Ueltschi for useful discussions. 
The research of T.B. was partly supported by ANR
project RMTQIT (Grant No. ANR-12-IS01-0001-01) and by 
ANR contract ANR-14-CE25-0003-0.
The research of V.J. was partly supported by NSERC. 
Y.P. was partly supported by ANR contract ANR-14-CE25-0003-0. Y.P. also wishes to thank UMI-CRM for financial support and McGill University for its hospitality.
The work of C.-A.P. has been
carried out in the framework of the Labex Archim\`ede (ANR-11-LABX-0033) and of
the A*MIDEX project (ANR-11-IDEX-0001-02), funded by the ``Investissements
d'Avenir'' French Government program managed by the French National Research
Agency (ANR).

\section{Results}

\subsection{Level I: Entropies}
\label{sec-entropies}

Let $\cX$ be a finite set. We denote by $\cP_\cX$ the set of all probability measures on 
$\cX$. The Gibbs-Shannon entropy\footnote{In the sequel we will just refer to it as
{\em the entropy}.} of $P\in\cP_\cX$ is
\[
S(P)=-\sum_{x\in\cX}P(x)\log P(x).
\] 
The map $\cP_\cX\ni P\mapsto S(P)$  is continuous and  takes values in $[0,\log|\chi|]$. The entropy is subadditive: 
if $\cX=\cX_1\times\cX_2$ and $P_1, P_2$ are the respective marginals of $P\in\cP_\cX$, 
then 
\beq
S(P)\leq S(P_1)+S(P_2),
\label{Ssubadd}
\eeq
with the  equality iff $P=P_1\times P_2$. The entropy is also concave and almost convex: 
if $P_k\in\cP_\cX$ for $k=1,\ldots,n$ and $p_k\geq 0$ are such that $\sum_{k=1}^n p_k=1$, 
then 
\beq
\sum_{k=1}^n p_kS(P_k)
\leq S\left(\sum_{k=1}^n p_kP_k\right)
\leq S(p_1,\ldots,p_n)+ \sum_{k=1}^n p_kS(P_k),
\label{tri-que}
\eeq
where $S(p_1,\ldots,p_n)=-\sum_{k=1}^n p_k\log p_k$. 

The set $\supp\,P=\{ x\in\cX\,|\,P(x)\not=0\}$ is called the support of $P\in\cP_\cX$.
For $\alpha\in\rr$, the R\'enyi $\alpha$-entropy of $P$ is defined by
\[
S_\alpha(P)=\log\left[\sum_{x\in\supp\,P} P(x)^\alpha\right].
\]
The map $\alpha\mapsto S_\alpha(P)$ is real analytic and convex. Obviously, 
\[
\left.\frac{\d\ }{\d\alpha}S_\alpha(P)\right|_{\alpha=1}=-S(P).
\]
The relative entropy of the pair $(P,Q)\in\cP_\cX\times\cP_\cX$ is
\[
S(P|Q)=\begin{cases}
\ds\sum_{x\in\cX} P(x)\log\frac{P(x)}{Q(x)},&\mbox{if }\supp\,P\subset\supp\,Q;\\[6pt]
+\infty,&\mbox{otherwise.}
\end{cases}
\]
The map $(P,Q)\mapsto S(P|Q)$ is lower semicontinuous and jointly convex. One easily shows that 
$S(P|Q)\geq0$ with equality iff $P=Q$. As a simple consequence, we note the log-sum 
inequality,\footnote{We use the 
conventions $0/0=0$ and $x/0=\infty$ for $x>0$, $\log 0=-\infty$, 
$\log\infty=\infty$, $0\cdot(\pm\infty)=0$.}
\beq
\sum_{j=1}^M a_j\log\frac{a_j}{b_j}
\geq a\log\frac{a}{b},\qquad 
a=\sum_{j=1}^M a_j,\quad
b=\sum_{j=1}^M b_j,
\label{log-sum}
\eeq
valid for non-negative $a_j,b_j$.

If ${\cal Y}$ is another finite set, a matrix $[M(x, y)]_{(x,y)\in {\cal X}\times {\cal Y}}$ with non-negative entries is called stochastic if 
for all $x\in {\cal X}$, $\sum_{y \in {\cal Y}}M(x, y)=1$. A stochastic matrix induces a transformation $M: \cP_\cX\rightarrow 
\cP_{\cal Y}$ by 
\[
M(P)(y)=\sum_{x\in {\cal X}}P(x)M(x, y).
\]
The relative entropy is monotone with respect to stochastic transformations:
\beq
S(M(P)|M(Q))\leq S(P|Q).
\label{MSrel}
\eeq
For $\alpha\in\rr$, the R\'enyi relative $\alpha$-entropy of a pair $(P,Q)$ satisfying 
$\supp \,P= \supp \,Q$ is defined by 
\[
S_\alpha(P|Q)=\log\left[\sum_{x\in\cX}P(x)^{1-\alpha}Q(x)^{\alpha}\right].
\]
The map $\rr\ni\alpha\mapsto S_\alpha(P|Q)$ is real analytic and convex. It clearly satisfies
\beq
S_\alpha(P|Q)=S_{1-\alpha}(Q|P),
\label{Renyi-Sym}
\eeq
and in particular $S_0(P|Q)=S_1(P|Q)=0$. Convexity further implies that
$S_\alpha(P|Q)\leq0$ for $\alpha\in[0,1]$ and $S_\alpha(P|Q)\geq0$ for 
$\alpha\not\in[0,1]$. The Renyi relative entropy  relates to the relative entropy through
\beq
\left.\frac{\d\ }{\d\alpha}S_\alpha(P|Q)\right|_{\alpha=0}=-S(P|Q), \qquad 
\left.\frac{\d\ }{\d\alpha}S_\alpha(P|Q)\right|_{\alpha=1}= S(Q|P).
\label{RenDiff}
\eeq
The map $(P, Q)\rightarrow S_\alpha(P|Q)$ is continuous. For $\alpha \in [0,1]$ this map 
is jointly concave and 
\beq S_\alpha(P|Q)\leq S_\alpha(M(P)|M(Q)).
\label{SaM}
\eeq

We note also that if $\varphi:\cX\to\cX$ is a bijection,
then 
\beq
S(P\circ\varphi |Q\circ\varphi)=S(P|Q), \qquad S_\alpha(P\circ\varphi |Q\circ\varphi )=S_{\alpha}(P|Q).
\label{gott}
\eeq

All the above entropies can be characterized by a suitable variant of the Gibbs  
variational principle. We note in particular that for any $P\in\cP_\cX$ and any
function $f:\cX\to\rr$ one has
\beq
F_P(f):=\log\left(\sum_{x\in\cX}\e^{f(x)}P(x)\right)
=\max_{Q\in\cP_\cX}\left(\sum_{x\in\cX}f(x)Q(x)-S(Q|P)\right),
\label{relatHamiltonian}
\eeq
and that the maximum is achieved by the measure
$$
P_f(x)=\e^{f(x)-F_P(f)}P(x).
$$

For further information about these fundamental notions we refer the reader 
to~\cite{AD,OP}.

\subsection{Level I: Entropy production rate}
\label{sec-ep}

It follows from Eq.~\eqref{holds} that $\supp\,\P_T$ and $\supp\,\wP_T$ have 
the same cardinality. Thus, if either $\supp\,\P_T\subset\supp\,\wP_T$ or 
$\supp\,\wP_T\subset\supp\,\P_T$, then $\supp\,\P_T=\supp\,\wP_T$. 
\begin{quote}{\bf Notation.} 
In the sequel $\P_T^\#$ denotes either $\P_T$ or $\wP_T$.
\end{quote}
The relation 
\beq
\P_T^\#(\omega_1,\ldots,\omega_T)
=\sum_{\omega_{T+1}\in\cA}\P_{T+1}^\#(\omega_1,\ldots,\omega_T,\omega_{T+1})
\label{merry}
\eeq
further gives that if $\supp\,\P_T\not=\supp\,\wP_T$ for some $T$, then
$\supp\,\P_{T^\prime}\not=\supp\,\wP_{T^\prime}$ for all $T^\prime>T$. 

Define the function
$$
\Omega_T\ni\omega\mapsto\sigma_T(\omega)=\sigma_T(\omega_1,\ldots,\omega_T)
=\log\frac{\P_T(\omega_1,\ldots,\omega_T)}{\wP_T(\omega_1,\ldots,\omega_T)}.
$$
Note that $\sigma_T$ takes value in $[-\infty,\infty]$ and satisfies
$\sigma_T\circ\Theta_T=-\sigma_T$. The family of random variables  
$\{\sigma_T\}_{T\geq 1}$ quantifies the irreversibility, or equivalently, the entropy 
production of our measurement process. The  notion of entropy production of dynamical 
systems goes back to seminal works~\cite{ECM,ES,GC1,GC2}; 
see~\cite{JPR,Ku1,LS,Ma1,Ma2,MN,MV,RM}.

The expectation value of $\sigma_T$ w.r.t.\;$\P$ is well-defined and is equal to the 
relative entropy of the pair $(\P_T,\wP_T)$. More precisely, one has
\beq
\EE[\sigma_T]=\sum_{\omega\in\Omega_T}\sigma_T(\omega)\P_T(\omega)
=S(\P_T|\wP_T)=S(\P_T\circ \Theta_T|\wP_T\circ \Theta_T)=S(\wP_T|\P_T).
\label{sigLdef}
\eeq
The log-sum inequality~\eqref{log-sum} and Eq.~\eqref{merry} give the pointwise inequality
\[
\sum_{\omega_{T+1}\in\cA}\P_{T+1}(\omega_1,\ldots,\omega_{T+1})
\log\frac{\P_{T+1}(\omega_1,\ldots,\omega_{T+1})}{\wP_{T+1}(\omega_1,\ldots,\omega_{T+1})}
\geq\P_T(\omega_1,\ldots,\omega_T)
\log\frac{\P_T(\omega_1,\ldots,\omega_T)}{\wP_T(\omega_1,\ldots,\omega_T)},
\]
and summing over all $\omega\in\Omega_T$ gives
\beq
\EE[\sigma_{T+1}]\geq\EE[\sigma_T]\ge0.
\label{sigmon}
\eeq
Note that dividing the above pointwise inequality by $\P_T(\omega_1,\ldots,\omega_T)$
shows that $\sigma_T$ is a submartingale w.r.t.\;the natural filtration. The martingale
approach to the statistics of repeated measurement processes has been used in several
previous studies;  see~\cite{KM1,KM2,BB,BBB} and references therein. In this work, however, 
we shall base our investigations on subadditive ergodic theory which provides
another perspective on the subject.

The definition and the properties of the entropy production we have described so far are 
of course quite general, and are applicable to any $\phi$-invariant probability measure 
$\QQ$ on $(\Omega,\cF)$ with $\QQ_T$ being the marginal of $\QQ$ on $\Omega_T$ and 
$\widehat{\QQ}_T=\QQ_T\circ\Theta_T$. The remaining results of the present  and all  results of  next sections, 
however, rely critically on a subadditivity property  of $\P$ described in Lemma \ref{key-simple-1}. 

Since, in view of~\eqref{sigmon}, the 
cases where $\EE[\sigma_T]=\infty$ for some $T$ are of little interest, in what follows 
we shall assume:
\begin{quote}\label{B-def}
{\bf Assumption (B)} $\supp\,\P_T=\supp\,\wP_T$ for all $T\ge1$.
\end{quote}
\newcommand{\AssumptionB}{{\hyperref[B-def]{{\rm (B)}}}}%

We shall say that a positive map $\Psi:\cB(\cH)\to\cB(\cH)$ is strictly positive, and
write $\Psi>0$, whenever $\Psi[X]>0$ for all $X>0$. One easily sees that this condition
is equivalent to $\Psi[\one]>0$. Under Assumption~\AssumptionA, a simple criterion for 
the validity of Assumption~\AssumptionB{}  is that $\Phi_a>0$ for all $a\in\cA$. Indeed, these two 
conditions imply
$$
(\Phi_{\omega_1}\circ\cdots\circ\Phi_{\omega_T})[\one]>0,\qquad
\rho^{-\tfrac12}(\Phi_{\theta(\omega_1)}\circ\cdots\circ\Phi_{\theta(\omega_T)})
[\rho^{\tfrac12}\one\rho^{\tfrac12}]\rho^{-\tfrac12}>0,
$$
for all $T\ge1$ and all $\omega=(\omega_1,\ldots,\omega_T)\in\Omega_T$. It follows
from the canonical construction~\eqref{crooks} that $\supp\,\P_T=\supp\,\wP_T=\Omega_T$
for all $T\ge1$.

\bet\label{shannon1}
\begin{enumerate}[{\rm (1)}]
\item The (possibly infinite) limit
\[
\ep(\cJ,\rho):=\lim_{T\to\infty}\frac1T\EE[\sigma_T]
\]
exists. We call it the {\em mean entropy production rate} of the 
repeated measurement process $(\cJ,\rho)$.
\item One has
\[
\ep(\cJ,\rho)=\sup_{T\geq 1}\frac1T\left(\EE[\sigma_T]+\log\min\sp(\rho)\right)\ge0,
\]
where $\sp(\rho)$ denotes the spectrum of the initial state $\rho$.
\item For $\P$-a.e.\;$\omega\in\Omega$, the limit 
\beq
\lim_{T\to\infty}\frac1T\sigma_T(\omega)=\bar\sigma(\omega)
\label{barsigdef}
\eeq
exists. The random variable $\bar\sigma$ satisfies $\bar\sigma\circ\phi=\bar\sigma$ and
\[
\EE[\bar\sigma]=\ep(\cJ,\rho).
\]
Its negative part $\bar\sigma_-=\frac12(|\bar\sigma|-\bar\sigma)$ satisfies 
$\EE[\bar\sigma_-]<\infty$. Moreover, if $\ep(\cJ,\rho)<\infty$, then
\[
\lim_{T\to\infty}\EE\left[\left|\frac1T\sigma _T-\bar\sigma\right|\right]=0.
\]
The number $\bar\sigma(\omega)$ is the {\em entropy production rate} of the process
$(\cJ,\rho)$ along the trajectory $\omega$.
\end{enumerate}
\eet

\noindent
{\bf Remark 1.} If $\P$ is $\phi$-ergodic, then obviously
$\bar\sigma(\omega)=\ep(\cJ,\rho)$ for $\P$-a.e.\;$\omega\in\Omega$.

{\bf Remark 2.} Since $\sigma_T\circ\Theta_T=-\sigma_T$, Part~(1) implies
$$
\lim_{T\to\infty}\frac1T\widehat{\EE}[\sigma_T]
=\lim_{T\to\infty}\frac1T\EE[-\sigma_T]
=-\ep(\cJ,\rho).
$$
Part~(3) applied to the OR dynamical system $(\Omega,\widehat{\P},\phi)$ yields that the 
limit~\eqref{barsigdef} exists $\wP$-a.e.\;and satisfies 
$\widehat{\EE}[\bar\sigma]=-\ep(\cJ,\rho)$. 
Assuming that $\P$ is $\phi$-ergodic, we have either $\P=\wP$ and hence 
$\ep(\cJ,\rho)=0$, or $\P\perp\wP$ (i.e., $\P$ and $\wP$ are mutually singular).

\noindent
{\bf Remark 3.} The assumption $\ep(\cJ,\rho)<\infty$ in Part~(3) will  be essential 
for most of the forthcoming results. It is ensured if $\Phi_a>0$ for all $a\in\cA$.
Indeed,  the latter condition implies that
$\Phi_a[\one]\geq\epsilon\one$ for some $\epsilon>0$ and all $a\in\cA$. Since
\[
\begin{split}
\EE[\sigma_T]\le\EE[-\log\wP_T]&=-\sum_{\omega\in\Omega_T} \P_T(\omega)\log \wP_T(\omega) =
-\sum_{\omega\in\Omega_T}\wP_T(\omega)\log \P_T(\omega)\\[2mm]
&=-\sum_{\omega\in\Omega_T}\wP_T(\omega)
\log\tr\left(\rho(\Phi_{\omega_1}\circ\cdots\circ\Phi_{\omega_T})[\one]\right),
\end{split}
\]
it follows that in this case $\ep(\cJ,\rho)\leq-\log\epsilon$. For a perfect instrument 
$\Phi_a[X]=V_a X V_a^\ast$, $\Phi_a[\one]\geq\epsilon\one$ for some $\epsilon>0$ and all $a\in\cA$ iff all $V_a$'s are invertible.

{\bf Remark 4.} 
For $i=1,2,\ldots$, let $(\cJ_i,\rho_i)$ denote processes on a Hilbert space $\cH_i$ with 
instrument $\cJ_i=\{\Phi_{i,a}\}_{a\in\cA_i}$. Set $\Phi_i=\sum_{a\in\cA_i}\Phi_{i,a}$ and
assume that OR processes $(\widehat{\cJ}_i,\widehat{\rho}_i)$ are induced by involutions
$\theta_i$ on $\cA_i$. Denote by $\PP_{i,T}^\#$ the probabilities induced on $\cA_i^T$
by these processes. Basic operations on instruments and the resulting measurement 
processes have the following effects on entropy production.

{\em Product.} The process $(\cJ_1\otimes\cJ_2,\rho_1\otimes\rho_2)$ is defined on the 
Hilbert space $\cH_1\otimes\cH_2$ by the instrument 
$\cJ_1\otimes\cJ_2:=\{\Phi_{1,a_1}\otimes\Phi_{2,a_2}\}_{(a_1,a_2)\in\cA_1\times\cA_2}$ 
and its OR is induced by the involution $\theta(a_1,a_2)=(\theta_1(a_1),\theta_2(a_2))$.
The probabilities induced on $(\cA_1\times\cA_2)^T$ by these processes are easily seen to 
be $\PP_T^\#((a_1,b_1),\ldots,(a_T,b_T))
=\PP_{1,T}^\#(a_1,\ldots,a_T)\PP_{2,T}^\#(b_1,\ldots,b_T)$ and it follows from the
equality in Relation~\eqref{Ssubadd} and Eq.~\eqref{sigLdef} that 
\[
\ep(\cJ_1\otimes\cJ_2,\rho_1\otimes\rho_2)=\ep(\cJ_1,\rho_1)+\ep(\cJ_2,\rho_2).
\]

{\em Sums.} The process $\left(\cJ_1\oplus\cJ_2,\rho^{(\mu)}\right)$ is
defined on the Hilbert space $\cH_1\oplus\cH_2$ with the initial state
$\rho^{(\mu)}=\mu\rho_1\oplus(1-\mu)\rho_2$, $\mu\in]0,1[$, and the instrument 
$\cJ_1\oplus\cJ_2=\{\Phi_a\}_{a\in\cA_1\cup\cA_2}$ with
$$
\Phi_a[A]=\bigoplus_{i\in\{1,2\}}1_{\cA_i}(a) J_i\Phi_{i,a}[J_i^\ast AJ_i]J_i^\ast,
$$
where $1_{\cA_i}$ denotes the characteristic function of $\cA_i$ and $J_i$ the natural 
injection $\cH_i\hookrightarrow\cH_1\oplus\cH_2$. It follows that
$$
\Phi[A]=\bigoplus_{i\in\{1,2\}}J_i\Phi_i[J_i^\ast AJ_i]J_i^\ast,
$$
and in particular $\Phi[\one]=\one$ and $\Phi^\ast[\rho^{(\mu)}]=\rho^{(\mu)}$. 
Assuming that $\theta_1$ and $\theta_2$
coincide on $\cA_1\cap\cA_2$, an OR process is induced by the involution defined on
$\cA_1\cup\cA_2$ by $\theta(a)=\theta_i(a)$ for $a\in\cA_i$. The probabilities induced
by these processes are  the convex combinations
$$
\PP_T^\#=\mu\,\PP_{1,T}^\#+(1-\mu)\PP_{2,T}^\#,
$$
where $\PP_{i,T}^\#$ is interpreted as a probability on $\cA_1\cup\cA_2$. The joint 
convexity of relative entropy and Relation~\eqref{sigLdef} yield the inequality
\[
\ep\left(\cJ_1\oplus\cJ_2,\rho^{(\mu)}\right)
\le\mu\,\ep(\cJ_1,\rho_1)+(1-\mu)\ep(\cJ_2,\rho_2).
\]
Note that if $\PP_1\not=\PP_2$, then the sum of two measurement processes is never ergodic.
Two extreme cases are worth noticing:\footnote{The alphabets $\cA_i$ are immaterial and
only serve the purpose of labeling individual measurements $\Phi_{i,a}$, thus the
identification of elements of $\cA_1$ and $\cA_2$ in $\cA_1\cap\cA_2$ is purely 
conventional. We note, however, that this identification  affects our definition of the sum of two instruments.} 

(a) If $\cA_1\cap\cA_2=\emptyset$ (disjoint sum), then $\PP_{1,T}\perp\PP_{2,T}$ 
and $\PP_T(a_1,\ldots,a_T)=\mu_i\PP_{i,T}(a_1,\ldots,a_T)$ ($\mu_1=\mu$, 
$\mu_2=1-\mu$) for all $T\ge1$ provided $a_1\in\cA_i$, i.e., the outcome of the 
first measurement selects the distribution of the full history. It immediately 
follows that in this case
\[
\ep\left(\cJ_1\oplus\cJ_2,\rho^{(\mu)}\right)
=\mu\,\ep(\cJ_1,\rho_1)+(1-\mu)\ep(\cJ_2,\rho_2).
\]
(b) If $\cA_1=\cA_2=\cA$ then $\PP_{1,T}$ and $\PP_{2,T}$ can be equivalent for 
all $T\ge1$. However, this does not preclude that $\PP_1\perp\PP_2$. This is 
indeed the case if $\PP_1$ and $\PP_2$ are ergodic and distinct. Then, there 
exists two $\phi$-invariant subsets $\cO_i\subset\cA^\nn$ such that 
$\PP_i(\cO_j)=\delta_{ij}$, $\PP(\cO_1\cup\cO_2)=1$, and
$$
\lim_{T\to\infty}\frac1T\sum_{t=0}^{T-1}f\circ\phi^t(\omega)=\EE_i[f],
$$
for $\PP$-a.e.\;$\omega\in\cO_i$ and all $f\in L^1(\cA^\nn,\d\PP)$. Thus, in this 
case, the selection occurs asymptotically as $T\to\infty$.

The above operations and results extend in an obvious way to finitely many 
processes.

{\em Coarse graining.} We shall say that $\cJ_2$ is coarser than $\cJ_1$ and 
write $\cJ_2\succ\cJ_1$ whenever $\cH_1=\cH_2$ and there exists a stochastic matrix 
$[M_{a_1a_2}]_{(a_1,a_2)\in\cA_1\times\cA_2}$ such that 
$$
\Phi_{2,a_2}=\sum_{a_1\in\cA_1}M_{a_1a_2}\Phi_{1,a_1}
$$
for all $a_2\in\cA_2$. Note that in this case $\Phi_1=\Phi_2$. In particular, 
$(\cJ_2,\rho)$ satisfies Assumption~\AssumptionA{} iff $(\cJ_1,\rho)$ does.
If $M_{\theta_1(a_1)\theta_2(a_2)}=M_{a_1a_2}$ for all $(a_1,a_2)\in\cA_1\times\cA_2$
then $\cJ_2\succ\cJ_1$ is equivalent to $\widehat{\cJ}_2\succ\widehat{\cJ}_1$ and
the induced probability distributions are related by
$$
\PP_{2,T}^\#=M_T(\PP_{1,T}^\#)
$$ 
where $[M_{T,ab}]_{(a,b)\in\cA_1^T\times\cA_2^T}$ is the stochastic matrix defined by
$$
M_{T,ab}=\prod_{i=1}^T M_{a_ib_i}.
$$
It follows from Inequality~\eqref{MSrel} and Relation~\eqref{sigLdef} that
\[
\ep(\cJ_2,\rho)\leq\ep(\cJ_1,\rho).
\]

{\em Compositions.} Assuming that $\cH_1=\cH_2=\cH$, $\rho_1=\rho_2=\rho$, and $\Phi_{1,a_1}\circ\Phi_2=\Phi_2\circ\Phi_{1,a_1}$
for all $a_1\in\cA_1$, the composition $(\cJ_1\circ\cJ_2,\rho)$ is the process 
with the alphabet $\cA_1\times\cA_2$, involution
$\theta(a_1,a_2)=(\theta_1(a_1),\theta_2(a_2))$ and instrument
$\cJ_1\circ\cJ_2=\{\Phi_{1,a_1}\circ\Phi_{2,a_2}\}_{(a_1,a_2)\in\cA_1\times\cA_2}$.
Setting $M_{(a_1,a_2)b}=\delta_{a_1,b}$ one easily sees that 
$\widetilde{\cJ}_1\succ\cJ_1\circ\cJ_2$ where 
$\widetilde{\cJ}_1=\{\Phi_{1,a}\circ\Phi_2\}_{a\in\cA_1}$. Due to our commutation
assumption, the probabilities induced by $\widetilde{\cJ}_1$ coincide
with that of $\cJ_1$. It follows that
\[
\ep(\cJ_1,\rho)\leq\ep(\cJ_1\circ\cJ_2,\rho).
\]

{\em Limits.} Let $(\cJ_n,\rho_n)$, $n\in\nn$, and $(\cJ,\rho)$ be processes with the 
same Hilbert space $\cH$, alphabet $\cA$ and involution $\theta$.  We say that 
$\lim(\cJ_n,\rho_n)=(\cJ,\rho)$ if 
\beq
\lim_{n\to\infty}\left(\|\rho_n-\rho\|
+\sum_{a\in\cA}\|\Phi_{n,a}-\Phi_a\|\right)=0.
\label{conv}
\eeq
Part~(2) of Theorem~\ref{shannon1} gives 
\beq
\ep(\cJ_n,\rho_n)=\sup_{T\geq 1}\frac{S(\PP_{n,T}|\wP_{n,T})+\log\lambda_n}{T},
\label{fri}
\eeq
where $\lambda_n=\min\sp(\rho_n)$. This relation and the lower semicontinuity of the  
relative entropy give that for any $T$,  
\[
\liminf_{n\to\infty}\ep(\cJ_n,\rho_n)\geq\liminf_{n\to\infty}
\frac{S(\PP_{n,T}|\wP_{n,T})+\log\lambda_n}{T}\geq 
\frac{S(\P_T|\wP_T)+\log\lambda}{T},
\]
where $\lambda=\min\sp(\rho)$. We used that the convergence~\eqref{conv} implies 
$\lim_{n\to\infty}\wP_{n,T}^\#=\wP_T^\#$ and $\lim_{n\to\infty}\lambda_n=\lambda$.
Since~\eqref{fri} also holds for $(\cJ,\rho)$ and $(\P_T,\wP_T)$, we derive
\[
\liminf_{n\to\infty}\ep(\cJ_n,\rho_n)\geq\ep(\cJ,\rho).
\]

\bigskip
The final result of this section deals with the vanishing of the entropy production.
\bep\label{chris-det}
If $\P=\wP$, then $\ep(\cJ,\rho)=0$. Reciprocally, in cases where $\ep(\cJ,\rho)=0$ the 
following hold:
\begin{enumerate}[{\rm (1)}]
\item $\ds\lim_{T\to\infty}\EE[\sigma_T]=\sup_{T\geq 1}\EE[\sigma_T]<\infty$.
\item The measures $\P$ and $\wP$ are mutually absolutely continuous with finite relative entropy.
\item If $\P$ is $\phi$-ergodic, then $\P=\wP$.
\end{enumerate}
\eep

\subsection{Level I: Stein's error exponents}
\label{sec-Stein}

For $\epsilon\in]0,1[$ and $T\ge1$ set
\beq
s_T(\epsilon):=\min\left\{\wP_T(\cT)\,\big|\,
\cT\subset\Omega_T,\P_T(\cT^c)\leq\epsilon\right\},
\label{SteinT}
\eeq
where $\cT^c=\Omega_T\setminus\cT$. Since $\wP_T=\P_T\circ\Theta_T$, the right hand side 
of this expression is invariant under the exchange of $\P_T$ and $\wP_T$. The {\em Stein 
error exponents} of the pair $(\P,\wP)$ are defined by
$$
\ubar s(\epsilon)=\liminf_{T\to\infty}\frac1T\log s_T(\epsilon),\qquad
\bar s(\epsilon)=\limsup_{T\to\infty}\frac1T\log s_T(\epsilon).
$$
In Section~\ref{sec-hyp} we shall interpret these error exponents 
in the context of hypothesis testing of the arrow of time.
\bet\label{stein}
Suppose that $\P$ is $\phi$-ergodic. Then, for all $\epsilon\in]0,1[$, 
\[
\ubar s(\epsilon)=\bar s(\epsilon)=-\ep(\cJ,\rho).
\]
\eet
We finish  with two remarks.

{\bf Remark 1.} Theorem~\ref{stein} and its proof give that
\begin{align*}
\ubar s&=\inf\left\{\liminf_{T\to\infty}\frac1T\log\wP_T(\cT_T)\,\,\bigg|\,\,
\cT_T\subset\Omega_T\mbox{ for }T\ge1,\mbox{ and }
\lim_{T\to\infty}\P_T(\cT_T^c)=0\right\},\\[3mm]
\bar s&=\inf\left\{\limsup_{T\to\infty}\frac1T\log\wP_T(\cT_T)\,\,\bigg|\,\,
\cT_T\subset\Omega_T\mbox{ for }T\ge1,\mbox{ and }
\lim_{T\to\infty}\P_T(\cT_T^c)=0\right\},
\end{align*}
satisfy $\bar s=\ubar s=-\ep(\cJ,\rho)$; see the final remark in 
Section~\ref{steinproof}.

{\bf Remark 2.} Theorems~\ref{shannon1} and~\ref{stein} also hold whenever
Assumption~\AssumptionA{} is replaced by the  following two conditions:
\begin{enumerate}[(a)]
\item There exists a density matrix $\rhoi>0$ such that $\Phi^\ast[\rhoi]=\rhoi$.
\item $\rho>0$.
\end{enumerate}
In this case, $\ep(\cJ,\rho)$ does not depend on the choice of $\rho$, i.e.,
$\ep(\cJ,\rho)=\ep(\cJ,\rhoi)$ for all density matrices $\rho>0$. 
Note however that if 
$\Phi^\ast[\rho]\not=\rho$, then, except in trivial cases, ${\mathbb P}$ is not $\phi$-invariant and the family 
$\{\widehat {\PP}_T\}_{T\geq 1}$ does not define a probability measure on $\Omega$.

\subsection{Level II: Entropies}
\label{sec-entropy2}

We denote by $\cP$ the set of all probability measures on $(\Omega,\cF)$ and by 
$\cPi\subset\cP$ the subset of all $\phi$-invariant elements of $\cP$.
We endow $\cP$ with the topology of weak convergence which coincides with the
relative topology inherited from the weak-$\ast$ topology of the dual of the Banach 
space $C(\Omega)$ of continuous functions on $\Omega$. This topology is metrizable and 
makes $\cP$ a compact metric space and $\cPi$ a closed convex subset of $\cP$.
Moreover, $\cPi$ is a Choquet simplex whose extreme points are the $\phi$-ergodic
probability measures on $\Omega$; see~\cite[Section~A.5.6]{Ru1}.

For $\QQ\in\cP$, $\QQ_T$ denotes the marginal of $\QQ$ on $\Omega_T$. Reciprocally,
a sequence $\{\QQ_T\}_{T\ge1}$, with $\QQ_T\in\cP_{\Omega_T}$ defines a unique 
$\QQ\in\cP$ iff
$$
\sum_{\omega_T\in\cA}\QQ_T(\omega_1,\ldots,\omega_T)
=\QQ_{T-1}(\omega_1,\ldots,\omega_{T-1})
$$
for all $T>1$. Moreover, $\QQ\in\cPi$ iff, in addition,
$$
\sum_{\omega_1\in\cA}\QQ_T(\omega_1,\ldots,\omega_T)=\QQ_{T-1}(\omega_2,\ldots,\omega_T)
$$
for all $T>1$. It follows that to each $\QQ\in\cPi$ we can associate a time-reversed
measure $\widehat{\QQ}\in\cPi$ with the marginals $\widehat{\QQ}_T=\QQ_T\circ\Theta_T$.
The map $\QQ\mapsto\widehat{\QQ}$ defines an affine involution $\Theta$ of $\cPi$.
Clearly, the map $\Theta$ preserves the set of $\phi$-ergodic probability measures. In the 
following we associate to $\QQ\in\cPi$ the family of signed measures on $\Omega$
defined by
$$
\QQ^{(\alpha)}:=(1-\alpha)\QQ+\alpha\widehat{\QQ}.
$$
Note that ${\widehat \QQ}^{(\alpha)}=\QQ^{(1-\alpha)}$ and that  $\QQ^{(\alpha)}\in\cPi$ for $\alpha\in[0,1]$.

If $\QQ\in\cPi$, then the subadditivity~\eqref{Ssubadd} of entropy gives
$S(\QQ_{T+T'})\leq S(\QQ_T)+S(\QQ_{T'})$ for all $T,T'\geq 1$, and Fekete's Lemma 
(Lemma~\ref{lemma-fek} below) yields that
\beq
h_\phi(\QQ)=\lim_{T\to\infty}\frac1T S(\QQ_T)=\inf_{T\geq 1}\frac1T S(\QQ_T).
\label{KS-ent}
\eeq
By the Kolmogorov-Sinai theorem~\cite[Theorem~4.18]{Wa}, the number $h_\phi(\QQ)$ is the 
Kolmogorov-Sinai entropy of the left shift $\phi$ w.r.t.\;the probability measure 
$\QQ\in\cPi$ and lies in the interval $[0,\log\ell]$, where $\ell$ is the number of elements of the alphabet 
${\cal A}$.  The map $\cPi\ni\QQ\mapsto 
h_\phi(\QQ)$ is upper semicontinuous. It is also  affine (recall the concavity/convexity bound~\eqref{tri-que}), i.e., 
\[
h_\phi(\lambda\QQ_1+(1-\lambda)\QQ_2)=\lambda h_\phi(\QQ_1)+(1-\lambda)h_\phi(\QQ_2)
\]
for all $\QQ_1,\QQ_2\in\cPi$ and $\lambda\in[0,1]$. Moreover, since 
$S(\QQ_T)=S(\widehat{\QQ}_T)$, one has $h_\phi(\QQ^{(\alpha)})=h_\phi(\QQ)$
for all $\alpha\in[0,1]$.

For $\QQ_1,\QQ_2\in\cP$, we write $\QQ_1\ll\QQ_2$ whenever $\QQ_1$ is absolutely 
continuous w.r.t.\;$\QQ_2$. In this case, ${\d\QQ_1}/{\d\QQ_2}$ denotes
the Radon-Nikodym derivative of $\QQ_1$ w.r.t.\;$\QQ_2$. The relative entropy of the
pair $(\QQ_1,\QQ_2)$ is
\[
S(\QQ_1|\QQ_2)=
\begin{cases}
\ds\int\log\left(\frac{\d\QQ_1}{\d\QQ_2}\right)\d\QQ_1,&\mbox{if } \QQ_1\ll\QQ_2;\\[6pt]
+\infty,&\mbox{otherwise.}
\end{cases}
\]
The map $\cP\times\cP\ni(\QQ_1,\QQ_2)\mapsto S(\QQ_1|\QQ_2)$ is  lower semicontinuous.
Moreover $S(\QQ_1|\QQ_2)\geq 0$ with equality iff $\QQ_1=\QQ_2$. The proofs of these 
basic facts can be found in~\cite{El}. If $f$ is a measurable function on $(\Omega,\cF)$,
we shall denote its expectation w.r.t.\;$\QQ\in\cP$ by
$$
\QQ[f]=\int_{\Omega}f(\omega)\d\QQ(\omega)
$$
whenever the right hand side is well defined.

\subsection{Level II: R\'enyi's relative entropy and thermodynamic formalism on $[0,1]$}
\label{sec-thermo}
  
For $T\ge1$ and $\alpha\in\rr$, we adopt the shorthand 
\beq
e_T(\alpha):=S_\alpha(\P_T|\wP_T),
\label{Renyi-ad}
\eeq
and note that, up to a sign change of its argument,
$$
e_T(\alpha)=\log\EE\left[\e^{-\alpha\sigma_T}\right]
$$
is the cumulant generating function of $\sigma_T$. In order to obtain
interesting statistical information about the asymptotic behavior of the random
variable $\sigma_T$, we shall investigate the existence and smoothness
properties of the large-$T$ limit of the function $T^{-1}e_T(\alpha)$. To get a
rough picture of the limiting function, avoiding the more subtle question of its
existence, we first describe the basic properties of the function
$$
\rr\ni\alpha\mapsto\bar e(\alpha)
=\limsup_{T\to\infty}\frac1T e_T(\alpha)\in[-\infty,+\infty].
$$

According to the general properties of R\'enyi's relative entropy listed in
Section~\ref{sec-entropies}, the function $\rr\ni\alpha\mapsto e_T(\alpha)$
is real analytic, convex, vanishing for $\alpha\in\{0,1\}$, non-positive on the
interval $[0,1]$ and non-negative on its complement. The symmetry 
property~\eqref{Renyi-Sym} and the invariance property~\eqref{gott} further yield
\beq
e_T(\alpha)=e_T(1-\alpha).
\label{gott-1}
\eeq
Finally,  from Relations~\eqref{RenDiff} and~\eqref{sigLdef} we deduce
$e_T^\prime(1)=-e_T^\prime(0)=\EE[\sigma_T]$, which implies the lower bound
$$
e_T(\alpha)\ge\left(\left|\alpha-\tfrac12\right|-\tfrac12\right)\EE[\sigma_T].
$$
From the general properties of convex functions (we refer the reader to~\cite{Ro} for
details) we infer that $\bar e$ is a convex function vanishing for $\alpha\in\{0,1\}$. It 
is non-positive on the interval $[0,1]$ and non-negative on its complement. It satisfies 
the symmetry
\beq
\bar e(1-\alpha)=\bar e(\alpha),
\label{barGC}
\eeq
and the lower bound
$$
\bar e(\alpha)\ge\left(\left|\alpha-\tfrac12\right|-\tfrac12\right)\ep(\cJ,\rho)
\ge-\tfrac12\ep(\cJ,\rho).
$$
The following dichotomy holds: either $\bar e$ is a proper convex function, i.e.,
$\bar e(\alpha)>-\infty$ for all $\alpha$, or it is improper and takes the value $-\infty$
for some $\alpha\in]0,1[$. In the first case, which is ensured by the condition
$\ep(\cJ,\rho)<\infty$, there exists $\tfrac12\le\kappa\le\infty$ such that $\bar e$ is 
continuous on $\cI=]\tfrac12-\kappa,\tfrac12+\kappa[$ and takes the value $+\infty$ on the 
(possibly empty) complement of the closure of $\cI$. In the second case, $\ep(\cJ,\rho)=\infty$, 
$\bar e(\alpha)=-\infty$ for all $\alpha\in]0,1[$ and $\bar e(\alpha)=+\infty$ for all 
$\alpha\in\rr\setminus[0,1]$; see~\cite[Theorem~7.2]{Ro}. 

The first results in this section concerns the existence and the characterization of 
the large-$T$ limit of $T^{-1}e_T(\alpha)$ for $\alpha\in[0,1]$. To motivate our
approach, note that the variational principle~\eqref{relatHamiltonian} implies that
for all $T\ge1$ and $\alpha\in\rr$,
\beq
\begin{split}
\frac1Te_T(\alpha)
&=\frac1T\max_{\QQ\in\cP}\left(-\alpha\QQ[\sigma_T]-S(\QQ_T|\PP_T)\right)\\
&=\frac1T\max_{\QQ\in\cP}\left(
-\alpha\QQ[\log\P_T]+\alpha\QQ[\log\wP_T]-\QQ[\log\QQ_T]+\QQ[\log\P_T]\right)\\
&\ge\max_{\QQ\in\cPi}\left(\frac1T\QQ^{(\alpha)}[\log\P_T]-\frac1T\QQ[\log\QQ_T]\right),
\end{split}
\label{finitevolgibbs}
\eeq
which indicates that the large-$T$ limit of the functional appearing on the
right hand side of this expression may be connected to the limiting cumulant generating
function of $\sigma_T$.

\bet\label{th-1}
\begin{enumerate}[{\rm (1)}] 
\item For all $\alpha\in[0,1]$ the (possibly infinite) limit 
\beq
e(\alpha):=\lim_{T\to\infty}\frac1T e_T(\alpha)
\label{ealphalimform}
\eeq
exists, is non-positive, and satisfies $e(0)=e(1)=0$. The function $[0,1]\ni\alpha\mapsto e(\alpha)$ is convex
and satisfies the symmetry
\beq
e(\alpha)=e(1-\alpha).
\label{eGCform}
\eeq
We shall call $e(\alpha)$ the {\em entropic pressure} of the 
repeated measurement process $(\cJ,\rho)$.
\item The following alternative holds: either $e(\alpha)=-\infty$ for all 
$\alpha\in]0,1[$, or $e(\alpha)>-\infty$ for all $\alpha\in[0,1]$.
\item For any $\QQ\in\cPi$ the (possibly infinite) limit 
$$
\varsigma(\QQ)
:=\lim_{T\to\infty}\QQ\left[-\frac1T\log\P_T\right]
$$
exists and is non-negative. The map $\cPi\ni\QQ\mapsto\varsigma(\QQ)$ is 
affine, lower semicontinuous, and satisfies $\varsigma(\P)=h_\phi(\P)$.
\item For $\QQ\in\cPi$ we set 
$$
f(\QQ):=h_\phi(\QQ)-\varsigma(\QQ).
$$
The map $\cPi\ni\QQ\mapsto f(\QQ)$ is affine and upper semicontinuous.
It satisfies $f(\P)=0$ and
\[
f(\QQ^{(\alpha)})\le \bar e(\alpha)
\]
for all $\QQ\in\cPi$ and $\alpha\in\rr$.
\setcounter{enumisaved}{\value{enumi}}
\end{enumerate}

\medskip
In the remaining statements we assume that $\inf_{\alpha\in[0,1]}e(\alpha)>-\infty$.

\begin{enumerate}[{\rm (1)}]
\setcounter{enumi}{\value{enumisaved}}
\item For $\alpha\in[0,1]$ one has
$$
e(\alpha)=\sup_{\QQ\in\cPi}f(\QQ^{(\alpha)}),
$$
and the set 
\[
\cP_\eq(\alpha):=\left\{\QQ\in\cPi\,|\, e(\alpha)=f(\QQ^{(\alpha)})\right\}
\]
is a non-empty, convex, compact subset of $\cPi$. It is a Choquet simplex and a face
of $\cPi$. The extreme points of $\cP_\eq(\alpha)$ are $\phi$-ergodic. 
\item The function $\alpha\mapsto e(\alpha)$ admits a left/right derivative 
$\partial^\mp e(\alpha)$ at each $\alpha\in]0,1[$, and
\beq
\partial^-e(\alpha)
=\inf_{\QQ\in\cP_\eq(\alpha)}(\varsigma(\QQ)-\varsigma(\widehat{\QQ}))
\le\sup_{\QQ\in\cP_\eq(\alpha)}(\varsigma(\QQ)-\varsigma(\widehat{\QQ}))
=\partial^+e(\alpha).
\label{rega}
\eeq
\item The left/right derivative of $e(\alpha)$ also exists at $\alpha=0/1$ and 
\beq
\begin{split}
\partial^+e(0)
&=\sup_{\QQ\in\cP_\eq(0)}(\varsigma(\QQ)-\varsigma(\widehat{\QQ}))
\geq-\ep(\cJ,\rho),\\[2mm]
\partial^-e(1)
&=\inf_{\QQ\in\cP_\eq(1)}(\varsigma(\QQ)-\varsigma(\widehat{\QQ}))\leq\ep(\cJ,\rho).
\end{split}
\label{c-sss}
\eeq
\item If $\P$ is ergodic, then $\cP_\eq(0)=\{\P\}$, $\cP_\eq(1)=\{\wP\}$, and 
$\partial^-e(1)=-\partial^+e(0)=\ep(\cJ,\rho)$.
\end{enumerate}
\eet
{\bf Remark 1.} As already mentioned, the condition 
$\ep(\cJ,\rho)<\infty$ ensures that $e(\alpha)>-\infty$ for $\alpha\in[0,1]$.
Thus, Remark~3 in Section~\ref{sec-ep} provides a sufficient condition for 
the validity of Parts~(5)--(8).  More precisely, if $\Phi_a[\one]\geq\epsilon\one$ for some $\epsilon>0$ and all $a\in\cA$,  
then $e(\alpha)\geq \log \ell + \log \epsilon$ for all $\alpha \in [0,1]$.  The normalization  $\sum_{a\in {\cal A}}\Phi_a(\one)=
\one$ ensures that $\epsilon \leq \frac{1}{\ell}$. 

{\bf Remark 2.} The symmetry~\eqref{eGCform} implies that the involution $\Theta$ maps 
$\cP_\eq(\alpha)$ onto $\cP_\eq(1-\alpha)$. 

{\bf Remark 3.} Regarding (\ref{rega}), note that for any ${\mathbb Q} \in {\cal P}_\phi$ such that $\varsigma({\mathbb Q})$ and 
$\varsigma(\widehat{\mathbb Q})$ are finite, 
\[\varsigma({\mathbb Q})-\varsigma(\widehat{\mathbb Q})=-\lim_{T\rightarrow \infty}\frac{1}{T}{\mathbb Q}(\sigma_T).
\]

{\bf Remark 4.} 
Indicating by a subscript the dependence of the entropic pressure,
Remark~4 after Theorem~\ref{shannon1} extends as follows. The 
entropic pressure of the product of two instruments is easily seen to be
$$
e_{(\cJ_1\otimes\cJ_2,\rho_1\otimes\rho_2)}(\alpha)
=e_{(\cJ_1,\rho_1)}(\alpha)+e_{(\cJ_2,\rho_2)}(\alpha),
$$
while the joint concavity of R\'enyi entropy and Eq.~\eqref{Renyi-ad} yield the 
following inequality for general sums:
$$
e_{(\cJ_1\oplus\cJ_2,\mu\rho_1\oplus(1-\mu)\rho_2)}(\alpha)
\geq\mu\,e_{(\cJ_1,\rho_1)}(\alpha)+(1-\mu)e_{(\cJ_2,\rho_2)}(\alpha).
$$
In the special case of a disjoint sum the identity
$$
e_{(\cJ_1\oplus\cJ_2,\mu\rho_1\oplus(1-\mu)\rho_2)}(\alpha)
=\max\left(e_{(\cJ_1,\rho_1)}(\alpha),e_{(\cJ_2,\rho_2)}(\alpha)\right)
$$
holds. If $\cJ_2\succ\cJ_1$, then Inequality~\eqref{SaM} shows that
\[
e_{(\cJ_1,\rho)}(\alpha)\geq e_{(\cJ_2,\rho)}(\alpha).
\]
It follows that for compositions we have
$$
e_{(\cJ_1\circ\cJ_2,\rho)}(\alpha)\geq e_{(\cJ_1,\rho)}(\alpha).
$$
Finally, if $\lim(\cJ_n,\rho_n)=(\cJ,\rho)$, then
\beq
\limsup_{n\to\infty}e_{(\cJ_n,\rho_n)}(\alpha)\leq e_{(\cJ,\rho)}(\alpha).
\label{ness}
\eeq
To prove this inequality, one uses that 
\[
e_{(\cJ_n,\rho_n)}(\alpha)
=\inf_{T\geq 1}\frac{e_{T,(\cJ_n,\rho_n)}(\alpha)-\log\lambda_n}{T}
\]
(see the proof of  Part~(1) of Theorem~\ref{th-1}), and  argues in the  same 
way as in the proof of the respective part of Remark~4 after 
Theorem~\ref{shannon1}.

To achieve a better control of the fluctuations of the entropic functional $\sigma_T$ and 
to derive Chernoff and Hoeffding error exponents for the hypothesis testing of the arrow 
of time, we must improve Theorem~\ref{th-1} in two ways: (a) by obtaining more 
information on the smoothness of the entropic pressure $e(\alpha)$, which, in the language
of thermodynamics, amounts to investigating the (non-)existence of dynamical phase 
transitions; and (b) by extending our control of the limit~\eqref{ealphalimform} outside 
of the interval $[0,1]$. In the next two sections we  settle these goals.

\subsection{Differentiability on $]0, 1[$}
\label{sec-differ}

Theorems~\ref{shannon1}, \ref{stein}, \ref{th-1} and Proposition~\ref{chris-det} 
are very general results. They hold for any $\PP\in\cP_\phi$ as long as 
$\supp\,\PP_T=\supp\,\widehat{\PP}_T$ for all $T\geq 1$ and the following 
structural inequality holds\footnote{In the literature, the inequality  (\ref{sun-1}) is sometimes called  the upper 
quasi-Bernoulli property.}: for some $C>0$ and all $T, T^\prime\geq 1$,
\beq 
\PP_{T+T^\prime}\leq C\,\PP_T\, \PP_{T^\prime}\circ\phi^T.
\label{sun-1}
\eeq
Note that if~\eqref{sun-1} holds for $\PP$, then it automatically holds for 
$\widehat{\PP}$ with the same constant $C$. Although property~\eqref{sun-1} could 
be difficult to establish for generic dynamical systems, it always holds for the
systems associated with repeated measurement process satisfying the regularity 
assumptions~\AssumptionA\, and~\AssumptionB; see Lemma~\ref{key-simple-1}. This 
is the reason that, until this point, we did not need any additional assumptions 
on our model.
 
To proceed with our analysis  and establish smoothness of the entropic pressure 
on the interval $]0,1[$, we need  to complement the inequality (\ref{sun-1})  
with  a suitable lower bound.

To put our assumptions in perspective, we  start by recalling the notion of a 
Gibbs measure as introduced by Bowen~\cite{Bo1,Bo2}. A measure $\PP\in\cP_{\phi}$ 
is called Gibbs if there exists a H\"older continuous function $\varphi:\Omega\to\rr$ and a constant $C>0$ such that for all $T\geq 1$, 
\beq 
C^{-1}\e^{-\sum_{t=0}^{T-1}\varphi\circ \phi^t}
\leq\PP_T
\leq C\e^{-\sum_{t=0}^{T-1}\varphi\circ \phi^t}.
\label{gibbs}
\eeq
The thermodynamic formalism of Gibbs measures is well-understood and is easily 
adapted to the study of entropy production. Indeed, the first proof of the 
fluctuation relation/theorem was done in this setting~\cite{GC1,GC2}; for an 
exposition of the full theory and references we refer the reader to~\cite{JPR,MV}.
One easily shows that a Gibbs measure satisfies the lower and upper 
quasi-Bernoulli properties
$$
C^{-3}\PP_T\, \PP_{T^\prime}\circ\phi^T\le
\PP_{T+T'}
\le C^3\PP_T\, \PP_{T^\prime}\circ\phi^T.
$$
However, except in special cases, the measures $\PP$ arising in repeated 
measurement processes do not satisfy the above quasi-Bernoulli lower bound
and hence are not Gibbs. Although the study of the thermodynamical formalism for 
non-Gibbsian measures can be traced back to the celebrated program of 
Dobrushin~\cite{Do,DoS} (see the reviews~\cite{Fe,LN,VE} for additional 
information), the approach we adopt in this work was developed only relatively 
recently, and is called  the subadditive thermodynamic formalism;
see~\cite{BaL,BV,CFH,CZC,Fa,FS,Fe1,Fe2,Fe3,FL,FK,IY,KW}. In this approach, one 
assumes the upper bound~\eqref{sun-1} and, depending on a setting, 
an appropriate lower bound, while completely abandoning the Gibbs 
condition~\eqref{gibbs}. We shall proceed similarly, keeping in mind that in our 
case the upper bound is always satisfied, while an effective lower bound has to  be 
based on an assumption that is suited for study of entropy production 
and is natural in the context of repeated quantum measurement processes.

To formulate this assumption we introduce some additional notation. We denote by
$$
\Omega_{\rm fin}=\bigcup_{T\ge0}\Omega_T
$$
the set of finite words. For $\bomega\in\Omega_{\rm fin}$  
we set $|\bomega|=T$ and $\P(\bomega)=\P_T(\omega_1,\ldots,\omega_T)$
whenever $\bomega=(\omega_1,\ldots,\omega_T)$.\footnote{By convention 
$\bomega\in\Omega_0$ is the empty word, i.e., $|\bomega|=0$ and $\P(\bomega)=1$.}

\begin{quote}\label{C-def} {\bf Assumption (C)} There exists $\tau\ge0$ such that
\[
C_\tau=\inf_{(\bomega,\bnu)\in\Omega_{\rm fin}\times\Omega_{\rm fin}}\,
\max_{\bxi\in\Omega_{\rm fin}\atop |\bxi|\le\tau}
\frac{\P(\bomega\bxi\bnu)\wP(\bomega\bxi\bnu)}
{\P(\bomega)\P(\bnu)\wP(\bomega)\wP(\bnu)}>0.
\]
\end{quote}
\newcommand{\AssumptionC}{{\hyperref[C-def]{{\rm (C)}}}}%

One of the main results of this work is:
\bet\label{th-2}
Suppose that Assumption~\AssumptionC{} holds. Then, for all $\alpha\in]0,1[$, the set 
$\cP_\eq(\alpha)$ is a singleton. In particular, the function $]0,1[\ni\alpha\mapsto e(\alpha)$ is differentiable.
\eet

Although Assumption~\AssumptionC{} may look technical, it is a natural optimal 
condition under which the subadditive thermodynamic formalism gives that 
$\cP_\eq(\alpha)$ is a singleton for $\alpha \in ]0,1[$. Moreover, 
Theorem~\ref{th-2} and its proof extend to any $\PP\in\cP_\phi$ for which 
Assumptions~\AssumptionB{} and  \AssumptionC{} hold and which satisfies the 
bound~\eqref{sun-1}.

As we shall discuss in~\cite{BJPP1}, Assumption~\AssumptionC{} is typically 
easy to verify in  applications to concrete examples. The next two propositions 
give sufficient conditions for~\AssumptionC{} that can be expressed directly in 
terms of the instrument $\{\Phi_a\}_{a\in\cA}$.

\bep Suppose that there exists an OR process 
$(\widehat{\cJ},\widehat{\rho}\,)$ with instrument 
$\widehat{\cJ}=\{\widehat{\Phi}_a\}_{a\in\cA}$ such that the completely  positive map 
$\Psi:\cB(\cH\otimes\cH)\to\cB(\cH\otimes\cH)$ defined by 
\beq
\Psi=\sum_{a\in\cA}\Phi_a\otimes\widehat{\Phi}_a
\label{never}
\eeq
is irreducible.\footnote{Recall that  $\Psi$ is irreducible if 
$\Psi[P]\le\lambda P$ for some orthogonal
projection $P\in\cB(\cH\otimes \cH)$ and some $\lambda>0$ implies $P\in\{0,\one\}$.}
Then Assumption~\AssumptionC{} holds.  
\label{A-C-1}\eep
{\bf Remark 1.} One easily shows that $\Psi\leq\Phi\otimes\widehat{\Phi}$, from which one deduces that
Assumption~\AssumptionC{} implies the irreducibility of $\Phi$. By~Theorem~\ref{thm-km}, 
the latter condition, in turn, implies that $\P$ is $\phi$-ergodic. 

{\bf Remark 2.}  If $(\widehat {\cal J}, \widehat \rho\,)$ is the canonical OR process (\ref{crooks}) and one of the $\Phi_a$'s is irreducible, 
then Proposition \ref{A-C-1} applies and  Assumption~\AssumptionC{} holds.  Thus, given any process $({\cal J}, \rho)$, ${\cal J}=\{\Phi_a\}_{a\in {\cal A}}$, 
and  a completely positive unital irreducible map $\Xi$ satisfying 
$\Xi^\ast(\rho)=\rho$, the instrument ${\cal J}_\epsilon= \{(1-\epsilon)\Phi_a,  \epsilon\,\Xi\}_{a\in {\cal A}}$, where $\epsilon \in ]0,1[$,  together with its canonical OR instrument, satisfies Assumption~\AssumptionC{}. The parameter $\epsilon$ can be interpreted as the 
probability that at each time $t=1,2, \cdots,$ no measurement is made, or that  the measurement result is lost/not read.  If $e_\epsilon(\alpha)$ is the entropic pressure 
of $({\cal J}_\epsilon, \rho)$ and $e(\alpha)$ of $({\cal J}, \rho)$, then for all $\alpha \in [0,1]$, 
\[e_\epsilon(\alpha)\geq \log(1-\epsilon) + e(\alpha),\]
while (\ref{ness}) gives $\lim_{\epsilon \downarrow 0} e_\epsilon(\alpha)\leq e(\alpha)$.
Hence, 
\beq\lim_{\epsilon \downarrow 0}e_\epsilon(\alpha)=e(\alpha).
\label{sat-conv}
\eeq
If $\inf_{\alpha \in [0,1]}e(\alpha)>-\infty$, then the convexity gives that the convergence (\ref{sat-conv}) is uniform 
on $[0,1]$, and that  for $\alpha \in ]0,1[$, 
\beq \partial^-e(\alpha)\leq \liminf_{\epsilon \downarrow 0}e_\epsilon^\prime(\alpha)\leq 
\limsup_{\epsilon \downarrow 0}e_\epsilon^\prime(\alpha)\leq 
\partial^+e(\alpha),
\label{sat-conv1}
\eeq
while for $\alpha =0/1$, 
\beq 
\partial^+e(0)\geq  \limsup_{\epsilon \downarrow 0}e_\epsilon^\prime(0), \qquad 
\partial^-e(1) \leq \liminf_{\epsilon \downarrow 0}e_\epsilon^\prime(0).
\label{sat-conv2}
\eeq

Using the canonical 
OR process~\eqref{crooks}, and invoking~\cite[Theorem~2.1]{JPW}, we obtain the following
simple algebraic criterion for validity of Proposition \ref{A-C-1} and  Assumption~\AssumptionC.
\bep\label{PROP-CondC}
Let
$$
\Phi_a[X]=\sum_{k=1}^{K_a}V_{a,k}^\ast XV_{a,k}
$$
be a Kraus decomposition of the instrument $\cJ=\{\Phi_a\}_{a\in\cA}$ and define
$$
W_{a,j,k}=V_{a,j}\otimes V^\ast_{\theta(a),k}.
$$
If the family $\{W_{a,j,k}\,|\,a\in\cA, j\in\llbracket1,K_a\rrbracket, 
k\in\llbracket1,K_{\theta(a)}\rrbracket\}$ acts 
irreducibly on $\cH\otimes\cH$, i.e., if the only subspaces of $\cH\otimes\cH$ which are 
invariant under all $W_{a,j,k}$ are $\{0\}$ and $\cH\otimes\cH$ itself, then the map 
(\ref{never}) is irreducible and Assumption~\AssumptionC{} holds.
\eep

\subsection{Full thermodynamic formalism}

To the best of our knowledge, Assumption~\AssumptionC{} is not sufficient to extend the 
thermodynamic formalism of Theorems~\ref{th-1} and \ref{th-2} to all $\alpha\in\rr$.
To  deal with this point we  strengthen~\AssumptionC{} as follows:

\begin{quote}\label{D-def} {\bf Assumption (D)} 
\[
D_0=\inf_{(\bomega,\bnu)\in\Omega_{\rm fin}\times\Omega_{\rm fin}}\,
\frac{\P(\bomega\bnu)}
{\P(\bomega)\P(\bnu)}>0.
\]
\end{quote}
\newcommand{\AssumptionD}{{\hyperref[D-def]{{\rm(D)}}}}%
Note that if~\AssumptionD{} holds for ${\mathbb P}$, then it also holds for $\widehat {\mathbb P}$ with the same constant $D_0$. 

\bet\label{th-3}
Suppose that Assumption~\AssumptionD{} holds. Then:
\begin{enumerate}[{\rm (1)}]
\item The limit 
\[
e(\alpha):=\lim_{T\to\infty}\frac1T e_T(\alpha)
\]
exists for all $\alpha\in\rr$, and the function $\rr\ni\alpha\mapsto e(\alpha)$ 
is differentiable. 
\item For any $\alpha\in\rr$, there exists a unique $\QQ_\alpha\in\cPi$ such that
\[
e(\alpha)=f((1-\alpha)\QQ_\alpha+\alpha\widehat{\QQ}_\alpha)
=\sup_{\QQ\in\cPi}f((1-\alpha)\QQ+\alpha\widehat{\QQ})
\]
Moreover, $\QQ_\alpha$ is $\phi$-ergodic and 
\[
e^{\prime}(\alpha)=\varsigma(\QQ_\alpha)-\varsigma(\widehat{\QQ}_\alpha).
\]
\end{enumerate}
\eet 

In our setting Assumption~\AssumptionD{} plays the role of the uniform 
hyperbolicity assumption in dynamical system theories and replaces/generalizes 
the Gibbs condition~\eqref{gibbs}. As we shall see in Section~\ref{secsec-FT}, 
Theorem~\ref{th-1} yields a global Fluctuation Theorem for repeated quantum 
measurement process. All known examples for which a global Fluctuation Theorem 
is proven are uniformly hyperbolic in a suitable sense. We believe that~\AssumptionD{} is the optimal general assumption for validity 
of Theorem~\ref{th-1} and the global Fluctuation Theorem.
 
Theorem~\ref{th-2}  and its proof extend to any $\PP\in\cP_\phi$ for which 
Assumptions~\AssumptionB{}, \AssumptionD{} hold and which satisfies the 
bound~\eqref{sun-1}. The next proposition gives sufficient condition
for~\AssumptionD{} in terms of the instrument $\{\Phi_a\}_{a\in\cA}$.

\bep Suppose that the map $\Phi_a$ is positivity improving\footnote{A 
positive map $\Psi:\cB(\cH)\to\cB(\cH)$ is positivity improving if $\Psi[X]>0$ for 
all $X\geq 0$.} for all $a\in\cA$. Then ~\AssumptionD{} holds.
\label{never-1}
\eep
{\bf Remark 1.} Given any process $({\cal J}, \rho)$, ${\cal J}=\{\Phi_a\}_{a\in {\cal A}}$, and unital completely positive positivity 
improving maps $\{\Psi_a\}_{a\in {\cal A}}$ satisfying $\Psi_a^\ast(\rho)=\rho$, 
the instrument 
\[{\cal J}_\epsilon= \left\{(1-\epsilon)\Phi_a +   \frac{\epsilon}{\ell}\Psi_a\right\}_{a\in {\cal A}},\]
where $\epsilon \in ]0,1[$,  satisfies  the assumption of the proposition. The deformation ${\cal J}_\epsilon$ of the original instrument ${\cal J}$ can be interpreted as an  effect of a  "noise" inherent in the measurement process. One easily verifies that for all $\alpha \in \rr$, $\liminf_{\epsilon \downarrow 0}e_\epsilon(\alpha) \geq \bar e(\alpha)$, while  (\ref{sat-conv})-(\ref{sat-conv2}) 
remain valid for $\alpha \in [0,1]$.

{\bf Remark 2.}  It is likely that under the assumption of the last proposition the function $\alpha \mapsto e(\alpha)$ is real analytic. The proof 
of such a result would require an adaptation of the transfer operator techniques to our settting \cite{Ba}. This point will be further 
discussed in \cite{BJPP3}.

\subsection{Level II: Large Deviations}
\label{SEC-LDP}

In this section we use  Theorems~\ref{th-1}, \ref{th-2} 
and~\ref{th-3} to study fluctuations of  the entropy production functional 
$\sigma_T$.

\subsubsection{Basic large deviations estimates}
\label{SEC-BasicLDP}

Assuming only Conditions~\AssumptionA--\AssumptionB{} and 
$\inf_{\alpha\in[0,1]}e(\alpha)>-\infty$, the 
following variant of the Large Deviation Principle follows from Theorem~\ref{th-1}\footnote{Note that $\bar e(\alpha)=e(\alpha)$ for $\alpha\in[0,1]$.}.

Recall that $\cI=]\tfrac12-\kappa,\tfrac12+\kappa[$ is the interior of the essential domain $\{\alpha\in\rr\,|\,\bar e(\alpha)<\infty\}$. Set
$$
s_\pm=-\partial^\pm\bar e(\tfrac12\mp\kappa),
$$
and note that
$$
-\infty\le s_-\le-\ep(\cJ,\rho)\le\ep(\cJ,\rho)\le s_+\le+\infty.
$$
For $s\in\rr$, set 
$$
I(s)=\sup_{\alpha\in\rr}(\alpha s-\bar e(-\alpha))=-\inf_{\alpha\in\rr}(\alpha s+\bar e(\alpha)).
$$
The function $\rr\ni s\mapsto I(s)$ is convex, finite and non-negative. It vanishes at 
$s=\ep(\cJ,\rho)$ and is non-increasing (resp.\;non-decreasing) for 
$s<\ep(\cJ,\rho)$ (resp. $s>\ep(\cJ,\rho)$). It satisfies
\beq
I(-s)-I(s)=s
\label{sand}
\eeq
as a consequence of the symmetry~\eqref{barGC}. By well-known properties of the 
Fenchel-Legendre transform,
\beq
-I(s)=
\begin{cases}
\bar e(\tfrac12+\kappa)+(\tfrac12+\kappa)s&\mbox{if }s\leq s_-;\\
\bar e(\alpha)+\alpha s&\mbox{if }s\in-\partial\bar e(\alpha)
\mbox{ for some }\alpha\in\cI;\\
\bar e(\tfrac12-\kappa)+(\tfrac12-\kappa)s&\mbox{if }s\geq s_+,
\end{cases}
\label{iloc1}
\eeq
where $\partial\bar e(\alpha)=[\partial^-\bar e(\alpha),\partial^+\bar e(\alpha)]$ denotes
the subdifferential of $\bar e$ at $\alpha$.

\bet Suppose that $\inf_{\alpha\in[0,1]}e(\alpha)>-\infty$. Then:
\begin{enumerate}[{\rm (1)}]
\item For any closed subset $C\subset\rr$ such that $\sup C<\infty$,
\[
\limsup_{T\to\infty}\frac1T\log\P_T\left(\left\{\omega\in\Omega_T\,\bigg|\,
\frac1T\sigma_T(\omega)\in C\right\}\right)\leq
-\inf_{s\in C}I(s).
\]
Moreover, the same estimate holds for all closed sets $C\subset\rr$ provided $\kappa>\tfrac12$.
\item If $s\in-\partial e(\alpha)$ for some $\alpha\in\,]0,1[$, 
then 
\[
\liminf_{T\to\infty}\frac1T\log\P_T\left(\left\{\omega\in\Omega_T\,\bigg|\,
\frac1T\sigma_T(\omega)<s\right\}\right)\geq-I(-\partial^+e(\alpha)).
\] 
\end{enumerate}
\label{ldp-1}
\eet
{\bf Remark.} This is a standard large deviation result~\cite{dH,DZ,El}\footnote{See 
also~\cite{JOPP,JOPS} for a pedagogical exposition.}, and we have stated it for reason 
of completeness. The same remark applies to Theorem~\ref{ldp-2} below.

\subsubsection{A local and a global Fluctuation Theorem}
\label{secsec-FT}

Theorems~\ref{th-2} and~\ref{th-3} allow us to refine the large deviations estimates
of the previous section and to obtain a full large deviations principle. 

\bet\label{ldp-2}
\begin{enumerate}[{\rm (1)}]
\item If Assumption~\AssumptionC{} holds, then for any open set 
$O\subset]-\ep(\cJ,\rho),\ep(\cJ,\rho)[$,
\beq
\lim_{T\to\infty}\frac1T\log\P_T\left(\left\{\omega\in\Omega_T\,\bigg|\, 
\frac1T\sigma_T(\omega)\in O\right\}\right)=- \inf_{s\in O} I(s).
\label{loc-LDP}
\eeq
\item Under Assumption~\AssumptionD, Relation~\eqref{loc-LDP} holds for any open set $O\subset\rr$.
\end{enumerate}
\eet

Parts~(1) and~(2) of Theorem~\ref{ldp-2} together with the relation~\eqref{sand} 
constitute, respectively, the local and global Fluctuation Theorem for our model; 
see~\cite{ECM,ES,GC1,GC2} for foundational works on the subject 
and~\cite{JPR,Ku1,LS,Ma1,Ma2,MN,MV,RM} for additional information.

{\bf Remark.} The following elementary observations provide a background for the 
Fluctuation Theorem. Denote by  $Q_T$ the law of the random variable 
$\frac1T\sigma_T$ w.r.t.\;$\P_T$: 
\[
Q_T(s)=\P_T\left(\left\{\omega\in\Omega_T\,|\,\sigma_T(\omega)=sT\right\}\right).
\]
Obviously, $Q_T(s)\not=0\Leftrightarrow Q_T(-s)\not=0$ and 
\beq
\int s\,\d Q_T(s)=\frac1T \EE[\sigma_T]\geq 0.
\label{jar-1}
\eeq
The relation~\eqref{gott-1} can be written as 
\[
\int\e^{-\alpha\sigma_T}\d\P_T=\int\e^{-(1-\alpha)\sigma_T}\d\P_T,
\]
and so for any $s$,
\beq
Q_T(-s)=\e^{-Ts}Q_T(s).
\label{jar-2}
\eeq
The relation~\eqref{jar-1} is the Jarzynski inequality in our setting. The normalization 
\[
1=\int \d \wP_T=\int\e^{-\sigma_T}\d\P_T=\int\e^{-s T} \d Q_T(s)
\]
is the Jarzynski identity, and~\eqref{jar-2} is the finite time Fluctuation Relation.  
Needless to say, the above elementary relations are  completely general and hold for any 
$\QQ\in\cPi$ and the associated entropy production observable. As emphasized 
in~\cite{GC1, GC2}, the mathematically and physically non-trivial aspects of the 
Fluctuation Theorem emerge through the Large Deviation Principle and the induced 
symmetry~\eqref{sand} of the rate function; see~\cite{JNPPS, JPS} for references and additional 
information regarding this point.

%%%%%%%%%%%%%%%
\subsection{Level II: Hypothesis testing} 
\label{sec-hyp}

By Remark~2 after Theorem~\ref{shannon1}, if $\P$ is $\phi$-ergodic and $\ep(\cJ,\rho)>0$,
then $\P$ and $\wP$ are mutually singular, i.e., concentrated on disjoint subsets of 
$\Omega$, whereas Assumption~\AssumptionB{} ensures that their marginals $\P_T$ and $\wP_T$ 
share a common support for all $T\ge1$. Hypothesis testing error exponents quantify the asymptotic 
separation of $\P_T$ and $\wP_T$ as $T\to\infty$. In our setting, given the physical 
interpretation of $\P_T$ and $\wP_T$ in terms of the process $(\cJ,\rho)$ and its outcome reversal 
$(\widehat{\cJ},\widehat{\rho}\,)$, one can say more colloquially that they substantiate the 
emergence of the arrow of  time in repeated quantum measurement processes.
 
The {\em Chernoff error exponents} of the pair $(\P,\wP)$ are defined by
$$
\ubar c=\liminf_{T\to\infty}\frac1T\log c_T,\qquad
\bar c=\limsup_{T\to\infty}\frac1T\log c_T,
$$
where
\[
c_T=\frac14\left(2-\sum_{\omega\in\Omega_T}\left|\P_T(\omega)-\wP_T(\omega)\right|\right).
\]
In the context of hypothesis testing, the number $c_T$ arises as follows. Let 
$(\widehat{\cJ},\widehat{\rho}\,)$ be an outcome reversal of $(\cJ,\rho)$. 
Consider the following two competing hypotheses:

\begin{description}
\item[Hypothesis I :] The observed quantum measurements are described by the
process $(\cJ,\rho)$.
\item[Hypothesis II :] The observed quantum measurements are described by  
the OR process $(\widehat{\cJ},\widehat{\rho}\,)$.
\end{description} 

By observing the first $T$ outcomes of the measurements we wish to determine with minimal 
error probability which of these two hypotheses is correct. More precisely, a 
{\em test} $\cT$ is a subset of $\Omega_T$ and its purpose is to discriminate between the 
two hypotheses. If the outcome $\omega$ of the first $T$ measurements is in $\cT$, one 
accepts~I and rejects~II. Otherwise, one accepts~II and rejects~I. To a given test $\cT$ 
one can associate two kinds of errors. A type-I error occurs when~I holds but 
$\omega\not\in\cT$. The conditional probability of such an error is $\P_T(\cT^c)$, where 
$\cT^c$ is the complement of $\cT$ in $\Omega_T$. If~II holds and $\omega\in\cT$, we get 
a type-II error, with conditional probability $\wP_T(\cT)$.

Assuming that the Bayesian probabilities assigned to the Hypothesis I and II are
$\tfrac12$\footnote{The discussion of the Chernoff error exponents easily extends to the case 
where these probabilities  are $p$ and $1-p$ for any $p\in]0,1[$.}, the total 
error probability is equal to $\tfrac12\P_T(\cT^c)+\tfrac12\wP_T(\cT)$
which we wish to minimize over $\cT$. The Neyman-Pearson lemma identifies the minimizer: 
if 
\beq
\ubar\cT_T=\{\omega\in\Omega_T\,|\,\P_T(\omega)\geq\wP_T(\omega)\},
\label{NPdef}
\eeq
then for any test $\cT\subset\Omega_T$,
\begin{align*}
\frac12\P_T(\cT^c)+\frac12\wP_T(\cT)&=\frac12-\frac12\left(
\P_T(\cT)-\wP_T(\cT)\right) 
\geq\frac12-\frac12\sum_{\omega\in\cT\cap\ubar \cT_T}
\left(\P_T(\omega)-\wP_T(\omega)\right)\\[6pt]
&=\frac12-\frac12\sum_{\omega\in\cT}
\left(\P_T(\omega)-\wP_T(\omega)\right)_+
\ge\frac12-\frac12\sum_{\omega\in\Omega_T}
\left(\P_T(\omega)-\wP_T(\omega)\right)_+\\[6pt]
&=\frac14\left(2-\sum_{\omega\in\Omega_T}\left|
\P_T(\omega)-\wP_T(\omega)\right|\right)=c_T.
\end{align*}
Observing that the two previous inequalities are saturated for 
$\cT=\ubar\cT_T$  one concludes that
\beq
\frac{1}{2}\P_T(\ubar \cT_T^c)+\frac{1}{2}\wP_T(\ubar\cT_T)=c_T.
\label{NPeq}
\eeq
Note also that the relation $\ubar\cT_T^c\subset \Theta_T(\ubar\cT_T)$ implies that $\wP_T(\ubar\cT_T)\geq \P_T(\ubar \cT_T^c)$.
\bet\label{thm-chernoff}
\begin{enumerate}[{\rm (1)}]
\item
 \[
\bar c\leq e\left(\tfrac12\right),
\]
and in particular $\bar c<0$ in cases where $\ep(\cJ,\rho)>0$.
\item If $\inf_{\alpha\in[0,1]}e(\alpha)>-\infty$, then
\[
\ubar c\geq e\left(\tfrac12\right)-\tfrac12\partial^+e\left(\tfrac12\right).
\]
\item If Assumption~\AssumptionC{} holds, then
$$
\ubar c=\bar c=\lim_{T\to\infty}\frac1T\log c_T=e(\tfrac12).
$$
\end{enumerate}
\eet
{\bf Remark. } The symmetry~\eqref{eGCform} implies that the convex function $e(\alpha)$ 
has a global minimum at $\alpha=\tfrac12$. Thus, if $e(\alpha)$ is differentiable at 
$\alpha=\tfrac12$, then $e^\prime(\tfrac12)=0$.
If $e(\alpha)$ is finite on  $]0,1[$, then it is also differentiable on 
$]0,1[$ outside a countable set, and one may expect that for a "generic" instrument 
one has $\partial^+e(\tfrac12)=e'(\tfrac12)=0$, in which case the conclusion of Part~(3) 
holds. 

\medskip
Theorem~\ref{thm-chernoff} provides a quantitative estimate for the emergence of the
arrow of time. Eq.~\eqref{NPeq} yields
\beq
\limsup_{T\to\infty}\frac1T\log\P_T(\ubar\cT_T^c)\leq \bar c, \qquad 
\limsup_{T\to\infty}\frac1T\log\wP_T(\ubar\cT_T)=\bar c.
\label{sun}
\eeq
Thus, if $\ep(\cJ,\rho)>0$, then the marginals $\P_T$ and $\wP_T$ respectively 
concentrate on the complementary subsets $\ubar\cT_T$ and $\ubar\cT_T^c$, with an 
exponential rate $\leq \bar c\le e(\tfrac12)<-\ep(\cJ,\rho)$. Note also  that if Assumption {\bf C} is satisfied, then 
\[\lim_{T\to\infty}\frac1T\log\P_T(\ubar\cT_T^c)=
\lim_{T\to\infty}\frac1T\log\wP_T(\ubar\cT_T)=e\left(\frac{1}{2}\right).
\]

\medskip
With the above interpretation of $\P_T(\cT_T^c)$ and $\wP_T(\cT_T)$ for 
$\cT_T\subset\Omega_T$, the number $s_T(\epsilon)$ introduced in Section~\ref{sec-Stein}
is the minimal probability of type-II errors that can be achieved by a test which ensures
that type-I errors have a maximal probability $\epsilon\in]0,1[$. Thus, the Stein error 
exponents control the exponential decay of type-II errors, 
$\e^{\ubar s(\epsilon)T}\lesssim\wP_T(\cT_T^c)\lesssim\e^{\bar s(\epsilon)T}$ 
as $T\to\infty$, in such tests.

The {\em Hoeffding error exponents} serve a similar purpose, but with a tighter 
constraint on the family $\{\cT_T\}_{T\ge1}$ of tests which are required to
ensure exponential decay of type-I errors with a minimal rate $s\ge0$. More
precisely,
\begin{align*}
\bar h(s)&=\inf_{\{\cT_T\}}\left\{\limsup_{T\to\infty}\frac1T\log\wP_T(\cT_T)\,\bigg|\, 
\limsup_{T\to\infty}\frac1T\log\P_T(\cT_T^c)<-s\right\},\\[3mm]
\ubar h(s)&=\inf_{\{\cT_T\}}\left\{\liminf_{T\to\infty}\frac1T\log\wP_T(\cT_T)\,\bigg|\,
\limsup_{T\to\infty}\frac1T\log\P_T(\cT_T^c)<-s\right\},\\[3mm]
h(s)&=\inf_{\{\cT_T\}}\left\{\lim_{T\to\infty}\frac1T\log\wP_T(\cT_T)\,\bigg|\, 
\limsup_{T\to\infty}\frac1T\log\P_T(\cT_T^c)<-s\right\},
\end{align*}
where in the last case the infimum is taken over all families of tests
for which  $\lim_{T\to\infty}\frac1T\log\wP_T(\cT_T)$ exists.  The Hoeffding error 
exponents satisfy $\ubar h(s)\leq \bar h(s)\leq h(s)$ and  have the same value 
if the roles of $\P$ and $\wP$ are exchanged.
Moreover, the functions $\ubar h(s)$, $\bar h(s)$, $h(s)$ are upper semicontinuous and 
right-continuous; see~\cite[Proposition~6.3]{JOPS}.

For $s\geq 0$ set
\[\psi(s)=-\sup_{\alpha\in [0,1[}\frac{-s\alpha -e(\alpha)}{1-\alpha}.
\]
If $e(\alpha)$ is finite on $[0,1]$, then $\psi(0)=-\partial^-e(1)=-\ep(\cJ,\rho)$. Moreover, 
$\psi$ is concave, increasing, and finite on $]0,\infty[$. 

\bet\label{thm-hoeffding}
Suppose that Assumption~\AssumptionC{} holds. Then for $s\geq 0$,  
\[
\ubar h(s)= \bar h(s)=h(s)=\psi(s).
\]
\eet

{\bf Remark.} This result follows from  Theorem 6.6 in \cite{JOPS}. The general arguments 
of \cite{JOPS} do not simplify in the special case considered here and, to avoid a 
complete repetition, we refer the reader to~\cite{JOPS}  for details of the proof. We 
also remark that the arguments of~\cite{JOPS} give that the estimate
$\bar h(s)\leq \psi(s)$ holds without any assumption on $e(\alpha)$. In analogy with 
Theorem~\ref{thm-chernoff}, one can also prove a suitable 
upper bound assuming only that $e(\alpha)$ is finite in $[0,1]$. We leave the details to 
interested reader.

%%%%%%%%%%%%%%%%%%%%%%%%%%%
%%%%%%%%%
\section{Level I: Proofs.}
\label{sec-ep-proofs}
%%%%%%%%%%%%

We start with some preliminaries. We first recall the well known
\bel\label{lemma-fek}
Let $\{a_t\}_{t\ge1}$ be a sequence of real numbers such that for some real number $c$ 
and all integers $t,s\geq 1$, 
\beq
a_{t+s}\leq a_t+a_s+c.
\label{fek-sub}
\eeq
Then, 
\[
\lim_{t\to\infty}\frac{a_t}{t}=\inf_{t\geq 1}\frac{a_t+c}{t}.
\]
\eel
{\bf Remark.} Lemma~\ref{lemma-fek} is a straightforward extension of the
classical Fekete lemma~\cite[Part~I, Chapter~3]{PS} which states that a subadditive 
sequence $b_{t+s}\le b_t+b_s$ satisfies
$$
\lim_{t\to\infty}\frac{b_t}t=\inf_{t\ge1}\frac{b_t}t.
$$
Indeed, it suffices to consider the sequence $b_t=a_t+c$.
Obviously, if the subadditivity assumption~\eqref{fek-sub} is replaced by 
super-additivity $a_{t+s}\geq a_t+a_s+c$, then
\[
\lim_{t\to\infty}\frac{a_t}{t}=\sup_{t\geq 1}\frac{a_t+c}{t}.
\]

\medskip
As already discussed in Section~\ref{sec-entropy2}, one celebrated application of 
Fekete's Lemma concerns the existence of the Kolmogorov-Sinai entropy. The
Shannon-McMillan-Breiman theorem is a deep refinement of this existence result.
\bet\label{SMBthm}
For $\QQ\in\cPi$, set $S_T(\omega )=-\log\QQ_T(\omega)$. Then the limit 
\[
s(\omega)=\lim_{T\to\infty}\frac1T S_T(\omega)
\]
exists and satisfies $s\circ\phi(\omega)=s(\omega)$ for $\QQ$-a.e.\;$\omega\in\Omega$.
Moreover, one has $\QQ[s]=h_\phi(\QQ)$ and 
\[
\lim_{T\to\infty}\QQ\left[\left|\frac1T S_T-s\right|\right]=0.
\]
\eet

The Kingman subadditive ergodic theorem is a deep refinement of the Fekete Lemma. 
\bet\label{kingman}
Let $\QQ\in\cPi$ and let $X_T:\Omega\to\rr$ be a sequence of random variables such  
that  $\QQ[X_1^+]<\infty$\footnote{$X_1^+=\max(X_1, 0).$}. Suppose further that for some 
real number $C$, all integers $T,T'\geq 1$, and $\QQ$-a.e.\;$\omega\in\Omega$,
$$
X_{T+T'}(\omega)\leq X_T(\omega)+X_{T'}\circ\phi^T(\omega)+C.
$$
Then, the limit
\[
\lim_{T\to\infty}\frac1T X_T(\omega)=x(\omega)
\]
exists and satisfies $x\circ\phi(\omega)=x(\omega)$ for $\QQ$-a.e.\;$\omega\in\Omega$.
Moreover, one has $\QQ[x^+]<\infty$, and
\[
\QQ[x]=\lim_{T\to\infty}\frac1T\QQ[X_T]\in[-\infty, \infty[.
\]
Finally, if $\QQ[|x|]<\infty$, then
\[
\lim_{T\to\infty}\QQ\left[\left|\frac1T X_T-x\right|\right]=0.
\]
\eet
\noindent
{\bf Remark.} The Shannon-McMillan-Breiman theorem cannot be directly deduced from 
Theorem~\ref{kingman}. However, an extension of the Kingman subadditive ergodic 
theorem due to Derriennic~\cite{Der} allows for such a deduction; see~\cite{BaY,Ja} 
for a pedagogical exposition of the proof. 

\medskip
The following subadditivity estimate plays a central role in our arguments. In what 
follows we  set $\lambda_0=\min\sp(\rho)$, so that $\lambda_0^{-1}\rho\geq\one$. 
Recall that $\P_T^\#$ denotes either $\P_T$ or $\wP_T$.
\bel\label{key-simple-1}
For all $T,T'\geq 1$, 
\[
\P_{T+T'}^\#\leq\lambda_0^{-1}\P_T^\#\,\P_{T'}^\#\circ\phi^T.
\]
\eel
\proof It suffices to consider the case $\P_T^\#=\P_T$.
Using the fact that $\tr(AB)\leq\tr(A)\|B\|$ for $A,B\ge0$, we can write
\begin{align*}
\P_{T+T'}(\omega_1,\ldots,\omega_{T+T'})
&=\tr(\rho(\Phi_{\omega_1}\circ\cdots\circ\Phi_{\omega_{T+T'}})[\one])\\[2mm]
&=\tr\left((\Phi_{\omega_T}^\ast\circ\cdots\circ\Phi_{\omega_1}^\ast)[\rho]
(\Phi_{\omega_{T+1}}\circ\cdots\circ\Phi_{\omega_{T+T'}})[\one]\right)\\[2mm]
&\leq\tr((\Phi_{\omega_T}^\ast\circ\cdots\circ\Phi_{\omega_1}^\ast)[\rho])\,
\|(\Phi_{\omega_{T+1}}\circ\cdots\circ\Phi_{\omega_{T+T'}})[\one]\|.
\end{align*}
The estimate
$$
\|(\Phi_{\omega_{T+1}}\circ\cdots\circ\Phi_{\omega_{T+T'}})[\one]\|
\le\tr\left((\Phi_{\omega_{T+1}}\circ\cdots\circ\Phi_{\omega_{T+T'}})[\one]\right)
\le\lambda_0^{-1}
\tr\left(\rho(\Phi_{\omega_{T+1}}\circ\cdots\circ\Phi_{\omega_{T+T'}})[\one]\right)
$$
thus leads to
\begin{align*}
\P_{T+T'}(\omega_1,\ldots,\omega_{T+T'})
&\leq\lambda_0^{-1}\tr(\rho(\Phi_{\omega_T}\circ\cdots\circ\Phi_{\omega_1})[\one])
\tr(\rho(\Phi_{\omega_{T+1}}\circ\cdots\circ\Phi_{\omega_{T+T'}})[\one])\\[2mm]
&=\lambda_0^{-1}\P_T(\omega_1,\ldots,\omega_T)\P_{T'}(\omega_{T+1},\ldots,\omega_{T+T'}).
\end{align*}
\qed

\subsection{Proof of Theorem~\ref{shannon1}}

\noindent
{\bf Parts (1--2)} Writing
\[
\EE[\sigma_T]=S(\P_T|\wP_T)
=-S(\P_T)-\sum_{\omega\in\Omega_T}\P_T(\omega)\log\wP_T(\omega),
\]
the subadditivity of entropy~\eqref{Ssubadd} and Lemma~\ref{key-simple-1} give 
\[
\EE[\sigma_{T+T'}]\geq\EE[\sigma_T]+\EE[\sigma_{T'}]+\log\lambda_0.
\]
Fekete's Lemma and the right inequality in~\eqref{sigmon} yield the results.

{\bf Part (3)} Note that 
$$
\sigma_T(\omega)=-X_T(\omega)-Y_T(\omega),
$$
where $X_T(\omega)=-\log\P_T(\omega)\ge0$ and  $Y_T(\omega)=\log\wP_T(\omega)\le0$.

Applying the Shannon-McMillan-Breiman theorem to $X_T$, we conclude that the limit
\beq
x(\omega)=\lim_{T\to\infty}\frac1T X_T(\omega)
\label{Xlimform}
\eeq
exists $\P$-a.s.\;and in $L^1(\Omega,\d\P)$, is $\phi$-invariant and non-negative.

Since $Y_T\le0$, and Lemma~\ref{key-simple-1} implies that 
$Y_{T+T'}\le Y_T+Y_{T'}\circ\phi^T-\log\lambda_0$, we can 
apply Kingman's subadditive ergodic theorem to conclude that the limit
\beq
y(\omega)=\lim_{T\to\infty}\frac1T Y_T(\omega)
\label{Ylimform0}
\eeq
exists $\P$-a.s., is $\phi$-invariant, non-positive, and satisfies
\beq
\EE[y]=\lim_{T\to\infty}\frac1T\EE[Y_T].
\label{Ylimform}
\eeq
It follows that~\eqref{barsigdef} holds $\P$-a.s.\;with $\bar\sigma=-x-y$.
Since both $x$ and $y$ are $\phi$-invariant, so is $\bar\sigma$.
The $L^1$-convergence in~\eqref{Xlimform} together with~\eqref{Ylimform} yield
that
$$
\EE[\bar\sigma]=\lim_{T\to\infty}\frac1T\EE[\sigma_T]=\ep(\cJ,\rho).
$$
From the fact that $\bar\sigma_-\le x$ we further deduce that 
$\EE[\bar\sigma_-]\le\EE[x]=h_\phi(\P)<\infty$. Finally, if $\ep(\cJ,\rho)<\infty$, 
then
$$
\EE[|\bar\sigma|]=\EE[\bar\sigma+2\bar\sigma_-]\le 2h_\phi(\P)+\ep(\cJ,\rho)<\infty,
$$
and Kingman's ergodic theorem  implies that~\eqref{Ylimform0}  holds in
$L^1(\Omega,\d\P)$. Thus, ~\eqref{barsigdef} also holds in $L^1(\Omega,\d\P)$.

\subsection{Proof of Proposition~\ref{chris-det}}

The first statement clearly follows from the fact that $\P=\wP$ implies $\sigma_T=0$.
For all $T\ge1$, set $\QQ_T^\#=\P_T^\#\times Q_T$ where $Q_T$ is an arbitrary probability 
measure on $\cA^{\rrbracket T,\infty\llbracket}$. Since $\QQ_T^\#\to\P^\#$ weakly as 
$T\to\infty$, the lower semicontinuity of the relative entropy gives
\beq
S(\P|\wP)\leq\liminf_{T\to\infty}S(\QQ_T|\widehat{\QQ}_T).
\label{RelEntEst}
\eeq
The obvious relation $S(\QQ_T|\widehat{\QQ}_T)=S(\P_T|\wP_T)$ and Eq.~\eqref{sigLdef}
further yield
\beq
S(\QQ_T|\widehat{\QQ}_T)=\EE[\sigma_T].
\label{sunnysaturday}
\eeq
In view of our assumption, Theorem~\ref{shannon1}~(2) writes
$$
\EE[\sigma_T]+\log\lambda_0\le \ep(\cJ,\rho)T=0,
$$
for all $T\ge1$. Thus, it follows from~\eqref{sigmon} that
$$
\lim_{T\to\infty}\EE[\sigma_T]=\sup_{T\ge1}\EE[\sigma_T]\le\log\lambda_0^{-1},
$$
which proves Part~(1). Combining the last estimate with~\eqref{RelEntEst} 
and~\eqref{sunnysaturday} we further get
$$
S(\P|\wP)\le\log\lambda_0^{-1}.
$$
In the same way one derives that $S(\wP|\P)\leq\log\lambda_0^{-1}$ and Part~(2) follows.
Finally, assuming that $\P$ is $\phi$-ergodic, Part~(3) follows from the facts that
$\wP\in\cPi$ and $\wP\ll\P$.

\subsection{Proof of Theorem~\ref{stein}} 
\label{steinproof}

For a given $\delta >0$ let $\cT_{T,\delta}=\left\{\omega\in\Omega_T\,|\,
\sigma_T(\omega)\geq c_\delta T\right\}$,
where $c_\delta=\delta$ whenever $\ep(\cJ,\rho)=\infty$ and
$c_\delta=\ep(\cJ,\rho)-\delta$ otherwise.
Since $\PP$ is $\phi$-ergodic, Theorem~\ref{shannon1}~(3) yields
\beq
\lim_{T\to\infty}\P_T(\cT_{T,\delta})=1.
\label{fla2}
\eeq
Thus, for $T$ large enough, $\P_T(\cT_{T,\delta}^c)\le\epsilon$ and consequently
$s_T(\epsilon)\le\wP_T(\cT_{T,\delta})$. We also have 
\beq
\wP_T(\cT_{T,\delta})=\wP_T\left(\left\{\omega\in\Omega_T\,|\,
1\le\e^{\sigma_T(\omega)-c_\delta T}\right\}\right)
\leq\e^{-c_\delta T}
\sum_{\omega \in\Omega_T}\e^{\sigma_T(\omega)}\wP_T(\omega)
=\e^{-c_\delta T}.
\label{fla3}
\eeq
Hence, for any $\delta>0$, 
\[
\limsup_{T\to\infty}\frac1T\log s_T(\epsilon)
\leq-c_\delta.
\]
Taking $\delta\downarrow0$ in the case $\ep(\cJ,\rho)<\infty$ and $\delta\uparrow\infty$
in the opposite case gives the upper bound 
\[
\limsup_{T\to\infty}\frac1T\log s_T(\epsilon)\leq-\ep(\cJ,\rho).
\]
To prove the  lower bound we may assume that $\ep(\cJ,\rho)<\infty$.
Let $\cU_{T,\epsilon}$ be a subset of $\Omega_T$ for which the minimum in
Eq.~\eqref{SteinT} is achieved, i.e., $s_T(\epsilon)=\wP_T(\cU_{T,\epsilon})$
and $\P_T(\cU_{T,\epsilon}^c)\le\epsilon$. For a given $\delta >0$ let
$
\cD_{T,\delta}=\left\{\omega\in\Omega_T\,|\,\sigma_T(\omega)
\leq(\ep(\cJ,\rho)+\delta)T\right\}
$.
Invoking Theorem~\ref{shannon1}~(3) again gives
\[
\lim_{T\to\infty}\P_T(\cD_{T,\delta})=1,
\]
and so, for $T$ large enough, $\P_T(\cD_{T,\delta}^c)\le\tfrac12(1-\epsilon)$. 
Since $\P_T(\cU_{T,\epsilon}^c)\le\epsilon$, we then have 
\begin{align*}
\tfrac12(1-\epsilon)&\le 1-(\P_T(\cU_{T,\epsilon}^c)+\P_T(\cD_{T,\delta}^c))\le
\P_T(\cU_{T,\epsilon}\cap\cD_{T,\delta})
=\int_{\cU_{T,\epsilon}\cap\cD_{T,\delta}}\e^{\sigma_T}\d\wP_T\\[4pt]
&\leq \e^{T(\ep(\cJ,\rho)+\delta)}\wP_T(\cU_{T,\epsilon}\cap\cD_{T,\delta})
\leq\e^{T(\ep(\cJ,\rho)+\delta)}\wP_T(\cU_{T,\epsilon}),
\end{align*}
and hence
$$
s_T(\epsilon)=\wP_T(\cU_{T,\epsilon})\ge\tfrac12(1-\epsilon)\e^{-T(\ep(\cJ,\rho)+\delta)}.
$$
It follows that for any $\delta >0$,
\[
\liminf_{T\to\infty}\frac1Ts_T(\epsilon)\geq-\ep(\cJ,\rho)-\delta,
\]
so that
\[
\liminf_{T\to\infty}\frac1Ts_T(\epsilon)\geq-\ep(\cJ,\rho).
\]
The result follows by combining the obtained lower and upper bounds.

\medskip\noindent
{\bf Remark.} The assertions of Remark~2 after Theorem~\ref{stein} can be deduced from the above arguments as follows.
Let $\{\cT_T\}_{T\ge1}$ be a sequence such that $\cT_T\subset\Omega_T$ and 
$\P_T(\cT_T^c)\to0$ as $T\to\infty$ and fix $\epsilon\in]0,1[$. For large enough $T$
one has $\P_T(\cT_T^c)\le\epsilon$ and hence $\wP_T(\cT_T)\ge s_T(\epsilon)$. It follows
from Theorem~\ref{stein} that
$$
-\ep(\cJ,\rho)\le\ubar s\le\bar s.
$$
Reciprocally, Eq.~\eqref{fla2} shows that the above tests $\cT_{T,\delta}$ are such that 
$\P_T(\cT_{T,\delta}^c)\to0$ as $T\to\infty$ and taking $\delta$ to $0$/$\infty$ in
Eq.~\eqref{fla3} as above yields $\bar s\le -\ep(\cJ,\rho)$.
%%%%%%%%%%%%%%%%%%%%%%%%%%%%%%%%%

%*****************************************************************************************
\section{Level II: Proofs.}
\label{sec-re-proofs}
%%%%%%%%%%%%
\subsection{Proof of Theorem~\ref{th-1}}

\noindent
{\bf Parts~(1)--(2)} By Definition~\eqref{Renyi-ad} and Lemma~\ref{key-simple-1}, we have
\begin{align*}
e_{T+T'}(\alpha)
&=\log\left(\sum_{\omega\in\Omega_{T+T'}}\P_{T+T'}(\omega)^{1-\alpha}
\wP_{T+T'}(\omega)^\alpha\right)\\
&\le\log\left(\sum_{\omega\in\Omega_{T+T'}}
\left(\lambda_0^{-1}\P_T(\omega)\P_{T'}\circ\phi^T(\omega)\right)^{1-\alpha}
\left(\lambda_0^{-1}\wP_T(\omega)\wP_{T'}\circ\phi^T(\omega)\right)^\alpha\right)\\
&=\log\left(\lambda_0^{-1}
\sum_{\omega\in\Omega_T}\P_T(\omega)^{1-\alpha}\wP_T(\omega)^\alpha
\sum_{\omega'\in\Omega_{T'}}\P_{T'}(\omega')^{1-\alpha}\wP_{T'}(\omega')^\alpha
\right)\\
&=e_T(\alpha)+e_{T'}(\alpha)+\log\lambda_0^{-1},
\end{align*}
for all $T,T'\ge1$ and all $\alpha\in[0,1]$. Thus, the existence of the limit $e(\alpha)$ 
is a consequence of the Fekete lemma which gives
\beq
e(\alpha)=\inf_{T\ge1}\frac1T\left(e_T(\alpha)+\log\lambda_0^{-1}\right).
\label{FekInf}
\eeq
The function $[0,1]\ni\alpha\mapsto e(\alpha)$ is convex as a limit of 
convex functions and upper semicontinuous as the infimum of the continuous functions.
The symmetry~\eqref{eGCform} and the values of $e(0)$ and $e(1)$ follow 
from the respective properties of $e_T(\alpha)$. Part~(2) follows from the
corresponding property of the function $\bar e$ established in the introductory 
discussion of Section~\ref{sec-thermo}.

\noindent
{\bf Part~(3)} By Lemma~\ref{key-simple-1}, one has
\beq
\QQ\left[-\log\P_{T+T'}\right]\ge\QQ\left[-\log\P_T\right]
+\QQ\left[-\log\P_{T'}\circ\phi^T\right]+\log\lambda_0.
\label{fpcpfc}
\eeq
The $\phi$-invariance of $\QQ$ and Fekete's lemma imply
$$
\varsigma(\QQ)=\lim_{T\to\infty}\QQ\left[-\frac1T\log\P_T\right]
=\sup_{T\ge1}\frac1T\left(\QQ[-\log\P_T]+\log\lambda_0\right).
$$
Since $\QQ\left[-\log\P_T\right]\ge0$, the limit $\varsigma(\QQ)$ takes value in  $[0, +\infty]$ (the limit is 
$+\infty$, for example,  in the cases where $\supp\,\QQ_T\not\subset\supp\,\P_T$ for some 
$T$). The function $\QQ\mapsto\varsigma(\QQ)$ is affine as a limit of affine functions and
lower semicontinuous as supremum of continuous functions. The last assertion
follows from Eq.~\eqref{KS-ent}.

{\bf Part (4)} Since the maps $\cPi\ni\QQ\mapsto h_\phi(\QQ)$ and
$\cPi\ni\QQ\mapsto-\varsigma(\QQ)$ are both affine and upper semicontinuous,
so is $\cPi\ni\QQ\mapsto f(\QQ)$. The last assertion of Part~(3) gives $f(\P)=0$.
From the variational inequality~\eqref{finitevolgibbs} 
we deduce that for all $T\ge1$, all $\alpha\in\rr$ and all $\QQ\in\cPi$,
$$
\frac1T e_T(\alpha)\ge\frac1T\QQ^{(\alpha)}[\log\P_T]-\frac1T\QQ[\log\QQ_T].
$$
Taking the limit $T\to\infty$ on both sides of this inequality yields the
last assertion.
{\tt }

\noindent
{\bf Part (5)} For $\alpha\in[0,1]$, the map $\cPi\ni\QQ\mapsto\QQ^{(\alpha)}\in\cPi$
is affine and continuous. Thus, we deduce from Part~(4) that 
$\cPi\ni\QQ\mapsto f(\QQ^{(\alpha)})$ is affine and upper semicontinuous. 
These two properties respectively imply that $\cP_\eq(\alpha)$ is convex and compact.
Let $m$ be a probability measure on $\cPi$ such that
$\int\QQ\,\d m(\QQ)=\bar\QQ\in\cP_\eq(\alpha)$. It follows that
$\bar\QQ^{(\alpha)}=\int\QQ^{(\alpha)}\d m(\QQ)$ and since $f$ is affine
and upper semicontinuous
$$
0=e(\alpha)-f(\bar\QQ^{(\alpha)})=\int(e(\alpha)-f(\QQ^{(\alpha)}))\d m(\QQ),
$$
from which we conclude that $m$ is concentrated on $\cP_\eq(\alpha)$, and hence that
$\cP_\eq(\alpha)$ is a face of $\cPi$. In particular, the extreme points of $\cP_\eq(\alpha)$
are $\phi$-ergodic; see~\cite[Section~A.5.6]{Ru1}.

It remains to prove that $\cP_\eq(\alpha)$ is non-empty. Since $\P^{(0)}=\P$ and $f(\P)=0=e(0)$, one has
$\P\in\cP_\eq(0)$. From the fact that $\QQ^{(1-\alpha)}=\widehat{\QQ}^{(\alpha)}$ we
further deduce that $\wP\in\cP_\eq(1)$. Thus, we may assume that $\alpha\in]0,1[$.

For $T\ge1$, the measure $Q_T\in\cP_{\Omega_T}$ defined by 
\[
Q_T(\omega)=\e^{-e_T(\alpha)}\P_T(\omega)^{1-\alpha}\wP_T(\omega)^\alpha
\]
achieves the maximum on the first line of~\eqref{finitevolgibbs}, 
\beq
e_T(\alpha)=S(Q_T)+(1-\alpha)Q_T[\log\P_T]+\alpha Q_T[\log \wP_T].
\label{san1}
\eeq
Decomposing $\Omega=\Omega_T^{\nn}$ into the product of blocks of size $T$,
we  extend $Q_T$ to a product probability measure 
$\QQ_{[T]}=Q_T^{\times\nn}\in\cP_{\phi^T}$. Setting 
\[
\QQ_{(T)}=\frac1T\sum_{t=0}^{T-1}\QQ_{[T]}\circ\phi^{-t},
\]
and observing that
$$
\QQ_{(T)}=\frac1{2T}\sum_{t=0}^{2T-1}\QQ_{[T]}\circ\phi^{-t},
$$
we obtain
$$
\QQ_{(T)}\circ\phi^{-1}=\frac1{2T}\sum_{t=1}^{2T}\QQ_{[T]}\circ\phi^{-t}
=\QQ_{(T)}-\frac1{2T}\left(\QQ_{[T]}-\QQ_{[T]}\circ\phi^{-2T}\right)=\QQ_{(T)},
$$
so that $\QQ_{(T)}\in\cPi$. Invoking a well known property of the Kolmogorov-Sinai entropy
(see, e.g., \cite[Theorem~4.13]{Wa}), its affine property and the Kolmogorov-Sinai
theorem, we can write
\beq
h_\phi(\QQ_{(T)})=\frac1T h_{\phi^T}(\QQ_{(T)})
=\frac1{T^2}\sum_{t=0}^{T-1}h_{\phi^T}(\QQ_{[T]}\circ\phi^{-t})
=\frac1{T^2}\sum_{t=0}^{T-1}\QQ_{[T]}[-\log Q_T\circ\phi^{t}].
\label{reallysick}
\eeq
For $t\in\llbracket0,T-1\rrbracket$, we derive a lower bound for 
$\QQ_{[T]}[-\log Q_T\circ\phi^{t}]$ in the following way (we denote by 
$\bomega\bnu\in\Omega_{T+T'}$ the word obtained by juxtaposition of the two words 
$\bomega\in\Omega_T$ and $\bnu\in\Omega_{T'}$)
\begin{align*}
\QQ_{[T]}[-\log Q_T\circ\phi^t]
&=-\sum_{{\bxi,\bnu\in\Omega_t}\atop{\bomega,\Beta\in\Omega_{T-t}}}
Q_T(\bxi\bomega)Q_T(\bnu\Beta)\log Q_T(\bomega\bnu)\\
&=-\frac12\sum_{{\bxi,\bnu\in\Omega_t}\atop{\bomega,\Beta\in\Omega_{T-t}}}
Q_T(\bxi\bomega)Q_T(\bnu\Beta)\log\left(Q_T(\bomega\bnu)Q_T(\Beta\bxi)\right)\\
&=\frac12\sum_{{\bxi,\bnu\in\Omega_t}\atop{\bomega,\Beta\in\Omega_{T-t}}}
Q_T(\bxi\bomega)Q_T(\bnu\Beta)\left[
\log\frac{Q_T(\bxi\bomega)Q_T(\bnu\Beta)}{Q_T(\bomega\bnu)Q_T(\Beta\bxi)}
-\log Q_T(\bxi\bomega)Q_T(\bnu\Beta)\right]\\
&\ge\frac12S(Q_T\times Q_T)=S(Q_T),
\end{align*}
where, in the last line, we have used the non-negativity of relative entropy.
Combining this lower bound with~\eqref{reallysick} we obtain
$$
h_\phi(\QQ_{(T)})\ge\frac1{T^2} \sum_{t=0}^{T-1}S(Q_T)=\frac1TS(Q_T),
$$
and~\eqref{san1} yields
\beq
\frac1T e_T(\alpha)\le h_\phi(\QQ_{(T)})
+(1-\alpha) Q_T\left[\frac1T\log\P_T\right]+\alpha Q_T\left[\frac1T\log \wP_T\right].
\label{flue}
\eeq

By the compactness of $\cPi$, there exists a sequence 
$T_k\uparrow\infty$ such that $\lim_{k\to\infty}\QQ_{(T_k)}=\QQ\in\cPi$. 
Applying Lemma~2.3 in~\cite{CFH} to the  sequence $f_T=\lambda_0^{-1}\P_T^\#$ 
we derive
\beq 
\limsup_{k\to\infty}Q_{T_k}\left[\frac{1}{T_k}\log\P_{T_k}^\#\right]
\leq\lim_{T\to\infty}\QQ\left[\frac1T\log\P_{T}^\#\right].
\label{san2}
\eeq
Combining these two inequalities with Relation~\eqref{flue} and Parts~(1) and~(3) 
gives
$$
e(\alpha)\le\limsup_{k\to\infty}h_\phi(\QQ_{(T_k)})-\varsigma(\QQ^{(\alpha)}).
$$
Finally, invoking the the upper semicontinuity of the Kolmogorov-Sinai entropy we derive
$$
e(\alpha)\le h_\phi(\QQ)-\varsigma(\QQ^{(\alpha)})=f(\QQ^{(\alpha)}),
$$
and so $\QQ\in\cP_\eq(\alpha)$.

\noindent
{\bf Part (6)}  We follow the proof of  \cite[Theorem~1.2]{Fe1} and  start by recalling the basic properties of the function
$[0,1]\ni\alpha\mapsto e(\alpha)$ which follow from the fact that it is convex and 
finite; see~\cite{Ro}.
\begin{enumerate}[(i)]
\item $\alpha\mapsto e(\alpha)$ is continuous.
\item For each $\alpha\in]0,1]$ the left derivative $\partial^-e(\alpha)$ exists.
\item For each $\alpha\in[0,1[$ the right derivative $\partial^+e(\alpha)$ exists.
\item $\partial^+e(0)\le\partial^-e(\alpha)\le\partial^+e(\alpha)\le
\partial^-e(\alpha')\le\partial^+e(\alpha')\le\partial^-e(1)$ for
$0\le\alpha\le\alpha'\le1$.
\item There is an at most countable subset $C\subset[0,1]$ such that
$\partial^-e(\alpha)=\partial^+e(\alpha)=e'(\alpha)$ holds for all 
$\alpha\in[0,1]\setminus C$.
\item For any $\alpha_0\in]0,1]$,
 $\lim_{\alpha\uparrow\alpha_0}\partial^+e(\alpha)
=\lim_{\alpha\uparrow\alpha_0}\partial^-e(\alpha)
=\partial^-e(\alpha_0)$. For any $\alpha_0\in[0,1[$,
 $\lim_{\alpha\downarrow\alpha_0}\partial^+e(\alpha)
=\lim_{\alpha\downarrow\alpha_0}\partial^-e(\alpha)
=\partial^+e(\alpha_0)$.
\een

Fix $\alpha\in]0,1[$. For any $\QQ\in\cP_\eq(\alpha)$ one has
$\varsigma(\QQ)<\infty$, and for $\epsilon>0$ small enough, 
\[
e(\alpha+\epsilon)
\geq f(\QQ^{(\alpha+\epsilon)})=f(\QQ^{(\alpha)})
+\epsilon(\varsigma(\QQ)-\varsigma(\widehat{\QQ}))
=e(\alpha)+\epsilon(\varsigma(\QQ)-\varsigma(\widehat{\QQ})).
\]
This gives that $\partial^+e(\alpha)\geq \varsigma(\QQ)-\varsigma(\widehat{\QQ})$, and 
so 
\beq 
\partial^+e(\alpha)\geq\sup_{\QQ\in\cP_\eq(\alpha)}
(\varsigma(\QQ)-\varsigma(\widehat{\QQ})).
\label{sick-tom}
\eeq
In the same way one derives 
\[
\partial^-e(\alpha)\leq\inf_{\QQ\in\cP_\eq(\alpha)}
(\varsigma(\QQ)-\varsigma(\widehat{\QQ})).
\]
In particular, if $e^\prime(\alpha)$  exists, then 
\beq
e^{\prime}(\alpha)=\varsigma(\QQ)-\varsigma(\widehat{\QQ})
\label{pain}
\eeq
for all $\QQ\in\cP_\eq(\alpha)$.

Let now $\alpha_k\in]0,1[\setminus C$ be a sequence such that 
$\alpha_k\downarrow\alpha$. By Properties~(v) and~(vi) $e^\prime(\alpha_k)$ exists, and 
\beq
\lim_{k\to\infty}e^\prime(\alpha_k)=\partial^+e(\alpha).
\label{headache}
\eeq
Choose 
$\QQ_k\in\cP_\eq(\alpha_k)$. Passing to a subsequence, which we also denote by 
$\alpha_k$, we may assume that $\lim_{k\to\infty}\QQ_k= \QQ\in\cP_\phi$.
Property~(i) and the upper semicontinuity of $f$ imply
\[
e(\alpha)=\lim_{k\to\infty}e(\alpha_k)
=\lim_{k\to\infty}f(\QQ_k^{(\alpha_k)})\leq f(\QQ^{(\alpha)}),
\]
and we conclude that $\QQ\in\cP_\eq(\alpha)$. The relation 
\[\lim_{k\to\infty}\left(h_\phi(\QQ_k^{(\alpha_k)})-\varsigma(\QQ_k^{(\alpha_k)})\right)= h_\phi(\QQ^{(\alpha)})-\varsigma(\QQ^{(\alpha)})
\]
together with the upper-semicontinuity  of $h_\phi$ and the  lower-semicontinuity of $\varsigma$ gives 
\[\liminf_{k\rightarrow \infty}\varsigma(\QQ_k^{(\alpha_k)})=\varsigma(\QQ^{(\alpha)}).\]
Passing again to a subsequence, 
we may assume that 
\beq \lim_{k\rightarrow \infty}\varsigma(\QQ_k^{(\alpha_k)})=\varsigma(\QQ^{(\alpha)}).
\label{sub-cold}
\eeq
The lower-semicontinuity of $\varsigma$ now gives that 
\[\liminf_{k\rightarrow \infty}\varsigma(\QQ_k)=\varsigma(\QQ), \qquad \liminf_{k\rightarrow \infty}\varsigma(\widehat \QQ_k)=\varsigma(\widehat \QQ).
\]
Passing again to a subsequence, we may assume that $\lim_{k\rightarrow \infty}\varsigma(\QQ_k)=\varsigma(\QQ)$. Then
(\ref{sub-cold}) gives that  along this final subsequence  $\lim_{k\rightarrow \infty}\varsigma(\widehat \QQ_k)=\varsigma(\widehat \QQ)$.
Combining this fact with  Relations~\eqref{pain} and~\eqref{headache} we derive
\[
\partial^+e(\alpha)=\lim_{k\to\infty}\varsigma(\QQ_k)-\varsigma(\widehat{\QQ}_k)=\varsigma(\QQ)-\varsigma(\widehat{\QQ})
\]
and so, by~\eqref{sick-tom},  
\[
\partial^+e(\alpha)=\sup_{\QQ\in\cP_\eq(\alpha)}
\varsigma(\QQ)-\varsigma(\widehat{\QQ}).
\]
An analogous argument yields 
\[
\partial^-e(\alpha)=\inf_{\QQ\in\cP_\eq(\alpha)}
\varsigma(\QQ)-\varsigma(\widehat{\QQ}).
\]

\noindent
{\bf Part (7)} The proof of relations~\eqref{c-sss} follows the proof of Part~(6). The 
inequalities follow from $\P\in\cP_\eq(0)$, $\wP\in\cP_\eq(1)$. The only difference is 
that for $\QQ\in\cP_\eq(0)$, $\varsigma(\QQ)<\infty$ while $\varsigma(\widehat\QQ)$ is allowed to 
take value $\infty$, which leads to the possibility that $\partial^+e(0)=-\infty$. The 
case $\QQ\in\cP_\eq(1)$ is analogous. 

\noindent
{\bf Part (8)} Assume that $\P$ is ergodic, let $\QQ\in\cP_\eq(0)$ and set
\[
a_T=S(\QQ_T|\P_T)=\QQ[-\log\P_T]-S(\QQ_T).
\]
Relation~\eqref{fpcpfc} and the subadditivity of entropy give that for all $T,T'\ge1$, 
\[
a_{T+T'}\geq a_T+ a_{T'}+\log\lambda_0.
\]
By Fekete's lemma, we have
\[
0=e(0)=f(\QQ)=\lim_{T\to\infty}\frac{a_T}{T}=\sup_{T\geq 1}\frac1T(a_T+\log\lambda_0).
\]
Hence, $S(\QQ_T|\P_T)\leq\log\lambda_0^{-1}$ for all $T\ge1$ and the lower semicontinuity 
of the relative entropy gives $S(\QQ|\P)\leq\log\lambda_0^{-1}$. This implies that 
$\QQ\ll\P$. Since $\QQ,\P\in\cP_\phi$, and $\P$ is ergodic, we have $\QQ=\P$. The proof 
of $\cP_\eq(1)=\{\wP\}$ is analogous.

%%%%%%%%%%%%%%
\subsection{Proof of Theorem \ref{th-2}}
\label{sec-theend}

We follow the strategy of~\cite{Fe3}.
In what follows  Assumption~\AssumptionC{} is supposed to hold without
further notice.

\bel\label{key-simple-3}
For any
$\bomega,\bnu\in\Omega_{\rm fin}$ there exists $\bxi\in\Omega_{\rm fin}$ such that
$|\bxi|\le\tau$ and
\[
\P(\bomega\bxi\bnu)^{1-\alpha}
\wP(\bomega\bxi\bnu)^{\alpha}\geq C_\tau\lambda_0\,
[\P(\bomega)\P(\bnu)]^{1-\alpha}
[\wP(\bomega)\wP(\bnu)]^{\alpha}
\]
for all $\alpha\in[0,1]$.
\eel
\proof Let $\bomega,\bnu\in\Omega_{\rm fin}$. Assumption~\AssumptionC{} gives
that for some $\bxi\in\Omega_t$ with $t\le\tau$,
\[
\frac{\P(\bomega\bxi\bnu)\wP(\bomega\bxi\bnu)}
{[\P(\bomega)\P(\bnu)]^{\alpha}[\wP(\bomega)\wP(\bnu)]^{1-\alpha}}
\geq C_\tau[\P(\bomega)\P(\bnu)]^{1-\alpha}
[\wP(\bomega)\wP(\bnu)]^{\alpha}.
\]
Lemma \ref{key-simple-1} gives
$$
\P^\#(\bomega\bxi\bnu)\leq\lambda_0^{-1}\P^\#(\bomega\bxi)\P^\#(\bnu)
\le\lambda_0^{-1}\sum_{\bxi'\in\Omega_t}\P^\#(\bomega\bxi')\P^\#(\bnu)
=\lambda_0^{-1}\P^\#(\bomega)\P^\#(\bnu).
$$
Hence, for $\alpha\in[0,1]$
$$
\P(\bomega\bxi\bnu)\wP(\bomega\bxi\bnu)
\le\P(\bomega\bxi\bnu)^{1-\alpha}\wP(\bomega\bxi\bnu)^\alpha
\lambda_0^{-1}[\P(\bomega)\P(\bnu)]^\alpha[\wP(\bomega)\wP(\bnu)]^{1-\alpha},
$$
and the result follows. \hfill\qed

\bel\label{lemma-long}
There exists a constant $C>0$ such that the following holds for all 
$\bomega\in\Omega_{\rm fin}$, all $T\ge1$ and all $\alpha\in[0,1]${\rm:}
\begin{enumerate}[\rm (1)]
\item
$\ds
\sum_{\bnu\in\Omega_T}
\P(\bomega\bnu)^{1-\alpha}\wP(\bomega\bnu)^{\alpha}
\geq C\e^{e_T(\alpha)}\P(\bomega)^{1-\alpha}\wP(\bomega)^{\alpha}.
$
\item
$\ds
\sum_{\bnu\in\Omega_T}
\P(\bnu\bomega)^{1-\alpha}\wP(\bnu\bomega)^{\alpha}
\geq C\e^{e_T(\alpha)}\P(\bomega)^{1-\alpha}\wP(\bomega)^{\alpha}.
$
\end{enumerate}
\eel
\proof We prove only Part~(1), the proof of Part~(2) is similar. In the case $\tau=0$ the results follow
immediately from Lemma~\ref{key-simple-3} with $C=C_0\lambda_0$. Consider now the case 
$\tau\ge1$. For any $\omega\in\Omega_{\rm fin}$ and any $\bxi\in\Omega_t$ with 
$t\le\tau$, we deduce from Lemma~\ref{key-simple-1}
\begin{align*}
\sum_{\nu\in\Omega_T}\P(\bomega\bxi\bnu)^{1-\alpha}\wP(\bomega\bxi\bnu)^\alpha
&\le\sum_{s=0}^\tau\sum_{\bnu\in\Omega_{T+s}}
\P(\bomega\bnu)^{1-\alpha}\wP(\bomega\bnu)^\alpha\\
&=\sum_{s=0}^\tau\sum_{\bnu\in\Omega_{T},\Beta\in\Omega_s}
\P(\bomega\bnu\Beta)^{1-\alpha}\wP(\bomega\bnu\Beta)^\alpha\\
&\le\lambda_0^{-1}\left(\sum_{s=0}^\tau\sum_{\Beta\in\Omega_s}
\P(\Beta)^{1-\alpha}\wP(\Beta)^\alpha\right)
\left(\sum_{\bnu\in\Omega_{T}}
\P(\bomega\bnu)^{1-\alpha}\wP(\bomega\bnu)^\alpha\right)\\
&=\left(\lambda_0^{-1}\sum_{s=0}^\tau\e^{e_s(\alpha)}\right)\sum_{\bnu\in\Omega_{T}}
\P(\bomega\bnu)^{1-\alpha}\wP(\bomega\bnu)^\alpha\\
&\le(\tau+1)\lambda_0^{-1}
\sum_{\bnu\in\Omega_{T}}\P(\bomega\bnu)^{1-\alpha}\wP(\bomega\bnu)^\alpha
\end{align*}
where, in the last line, we used the fact that $e_s(\alpha)\le0$ for $\alpha\in[0,1]$.
Since the above estimate holds for all $\bxi\in\Omega_{\rm fin}$ with $|\bxi|\le\tau$, Lemma \ref{key-simple-3} allows us to conclude that
\begin{align*}
\sum_{\bnu\in\Omega_{T}}\P(\bomega\bnu)^{1-\alpha}\wP(\bomega\bnu)^\alpha
&\ge\frac{\lambda_0}{\tau+1}\sum_{\nu\in\Omega_T}C_\tau\lambda_0
\P(\bnu)^{1-\alpha}\wP(\bnu)^\alpha
\P(\bomega)^{1-\alpha}\wP(\bomega)^\alpha\\
&=\frac{C_\tau\lambda_0^2}{\tau+1}
\e^{e_T(\alpha)}\P(\bomega)^{1-\alpha}\wP(\bomega)^\alpha.
\end{align*}
\hfill\qed

The super-additivity of the sequence $\{e_T(\alpha)\}_{T\ge1}$ for $\alpha\in[0,1]$ 
is an immediate consequence of the last lemma. Invoking Fekete's lemma thus yields
the following

\begin{corollary}\label{COR-super}
There exist a constant $c$ such that for all $\alpha\in[0,1]$ and all 
$T,T'\geq 1$,
\[
e_{T+T'}(\alpha)\geq e_T(\alpha)+e_{T'}(\alpha)+c.
\]
In particular,
\beq
e(\alpha)=\sup_{T\ge1}\frac1T(e_T(\alpha)+c)\in\,]-\infty,0]
\label{FekSup}
\eeq
for all $\alpha\in[0,1]$.
\end{corollary}

\bel\label{sss}
For any $\alpha\in]0, 1[$ there exists $\bar\QQ\in\cP_\phi$ such that, for some $C>0$, all $T>0$ and all 
$\bomega\in\Omega_T$,
\beq
C^{-1}\e^{-Te(\alpha)}\P(\bomega)^{1-\alpha}\wP(\bomega)^\alpha
\leq\bar\QQ(\bomega)
\leq C\e^{-Te(\alpha)}\P(\bomega)^{1-\alpha}\wP(\bomega)^\alpha.
\label{sick-0}
\eeq
Moreover, any such $\bar\QQ$ is $\phi$-ergodic.
\eel
\proof Throughout the proof, we fix $\alpha\in]0,1[$ and let $C$ denote a strictly 
positive constant which does not depend on $T$, but whose value may differ from 
place to place.

Combining sub- and super-additivity, we can find $c>0$ such that 
\beq
e_T(\alpha)+e_{T'}(\alpha)-c\leq e_{T+T'}(\alpha)\leq e_T(\alpha)+e_{T'}(\alpha)+c,
\label{seq-sub}
\eeq
for all $T,T'\geq 1$. For each $T\ge1$ let $\QQ_{[T]}\in\cP_{\phi^T}$ be defined as in 
the proof of Part~(5) of Theorem~\ref{th-1}, so that
\[
\QQ_{[T]}(\bomega)=\e^{-e_T(\alpha)}\P(\bomega)^{1-\alpha}\wP(\bomega)^\alpha
\]
for all $\bomega\in\Omega_T$. Let $\QQ$ be a limit point of the sequence 
$\{\QQ_{[T]}\}_{T\ge1}$, i.e., assume that for some sequence $T_k\uparrow\infty$
one has $\QQ=\lim_{k\to\infty}\QQ_{[T_k]}$. We claim that such a $\QQ$ satisfies the 
estimate~\eqref{sick-0}.

We first deal with the lower bound. Fix $T\geq 1$ and let $\bomega\in\Omega_T$. Then 
\beq
\QQ(\bomega)=\lim_{k\to\infty}\QQ_{[T_k]}(\bomega),
\label{mon-meet}
\eeq
and  Lemma~\ref{lemma-long} implies that for $T_k>T$,
$$
\QQ_{[T_k]}(\bomega)
=\e^{-e_{T_k}(\alpha)}\sum_{\bnu\in\Omega_{T_k-T}}
\P(\bomega\bnu)^{1-\alpha}\wP(\bomega\bnu)^{\alpha}
\geq C\e^{e_{T_k-T}(\alpha)-e_{T_k}(\alpha)}\P(\bomega)^{1-\alpha}\wP(\bomega)^\alpha.
$$
It follows from~\eqref{seq-sub} that 
$$
e_{T_k-T}(\alpha)-e_{T_k}(\alpha)\ge -e_T(\alpha)-c,
$$
while~\eqref{FekSup} allows us to write
\beq
e(\alpha)=\sup_{t\ge1}\frac1t(e_t(\alpha)+c)\geq\frac1T(e_T(\alpha)+c),
\label{lowe}
\eeq
and conclude that
\[
\QQ_{[T_k]}(\bomega)\geq C\e^{-Te(\alpha)}\P(\bomega)^{1-\alpha}\wP(\bomega)^\alpha.
\]
This inequality and~\eqref{mon-meet} yield that the measure $\QQ$ satisfies the lower 
bound in~\eqref{sick-0}.

We now turn to the upper bound. For $T_k>T$  and $\bomega\in\Omega_T$, 
Lemma~\ref{key-simple-1} gives 
$$
\QQ_{[T_k]}(\bomega)=
\e^{-e_{T_k}(\alpha)}\sum_{\bnu\in\Omega_{T_k-T}}
\P(\bomega\bnu)^{1-\alpha}\wP(\bomega\bnu)^\alpha
\leq C\e^{e_{T_k-T}(\alpha)-e_{T_k}(\alpha)}\P(\bomega)^{1-\alpha}\wP(\bomega)^\alpha.
$$
It follows from (\ref{seq-sub})  that 
$$
e_{T_k-T}(\alpha)-e_{T_k}(\alpha)\le -e_T(\alpha)+c,
$$
and~\eqref{FekInf} yields
\beq
e(\alpha)=\inf_{t\ge1}\frac1t(e_t(\alpha)-c)\leq\frac1T(e_T(\alpha)-c),
\label{upe}
\eeq
so that
\[
\QQ_{[T_k]}(\bomega)\leq C\e^{-Te(\alpha)}\P(\bomega)^{1-\alpha}\wP(\bomega)^\alpha.
\]
This inequality and~\eqref{mon-meet} yield that the upper bound in~\eqref{sick-0} holds 
for the measure $\QQ$. 

Note that the measure $\QQ$ needs not to be in $\cPi$. To deal with this point, we set
\[
\QQ_{(T)}=\frac1T\sum_{t=0}^{T-1}\QQ\circ\phi^{-t},
\]
and note that for $\bomega\in\Omega_{T'}$
\[
\QQ\circ\phi^{-t}(\bomega)=\sum_{\bnu\in\Omega_t}
\QQ(\bnu\bomega).
\]
The estimates~\eqref{sick-0} and 
Lemmas~\ref{key-simple-1}, \ref{lemma-long} lead to
\[
 C^{-1}\e^{e_t(\alpha)-(T'+t)e(\alpha)}\P(\bomega)^{1-\alpha}\wP(\bomega)^\alpha
\leq\QQ\circ\phi^{-t}(\bomega)
\leq C\e^{e_t(\alpha)-(T'+t)e(\alpha)}\P(\bomega)^{1-\alpha}\wP(\bomega)^{\alpha}.
\]
Since~\eqref{lowe} and~\eqref{upe} further imply $c\le e_t(\alpha)-te(\alpha)\le -c$,
we derive
$$
C^{-1}\e^{-T'e(\alpha)}\P(\bomega)^{1-\alpha}\wP(\bomega)^\alpha
\leq\QQ\circ\phi^{-t}(\bomega)
\leq C\e^{-T'e(\alpha)}\P(\bomega)^{1-\alpha}\wP(\bomega)^\alpha,
$$
for all $\bomega\in\Omega_{T'}$.
It follows that any limit point $\bar\QQ$ of the sequence $\{\QQ_{(T)}\}_{T\ge1}$,
which belongs to $\cPi$ by construction, also satisfies the estimates~\eqref{sick-0}.
 
Finally, we  proceed to show that any $\bar\QQ\in\cPi$ satisfying the 
estimates~\eqref{sick-0} is $\phi$-ergodic. To this end, let
$\cC_1$ and $\cC_2$ be two cylinder subsets of $\Omega$, more explicitely
$$
\cC_i=\{\omega\in\Omega\,|\,(\omega_1,\ldots,\omega_{r_i})\in C_i\},
$$
where $r_i>0$ and $C_i\subset\Omega_{r_i}$. For $t\ge r_1$, the lower bound 
in~\eqref{sick-0} yields
\begin{align*}
\bar\QQ(\cC_1\cap\phi^{-t}(\cC_2))
&=\sum_{\bomega_i\in C_i}\sum_{\bxi\in\Omega_{t-r_1}}
\bar\QQ(\bomega_1\bxi\bomega_2)\\
&\ge\sum_{\bomega_i\in C_i}\sum_{\bxi\in\Omega_{t-r_1}}C^{-1}\e^{-(t+r_2)e(\alpha)}
\P(\bomega_1\bxi\bomega_2)^{1-\alpha}
\wP(\bomega_1\bxi\bomega_2)^\alpha.
\end{align*}
Invoking Lemma~\ref{key-simple-3} and the fact that $e(\alpha)\le0$, we can write
\begin{align*}
\sum_{t=r_1}^{r_1+\tau}\bar\QQ(\cC_1\cap\phi^{-t}(\cC_2))
&\ge\sum_{\bomega_i\in C_i}\sum_{t=0}^\tau\e^{-te(\alpha)}
\sum_{\bxi\in\Omega_t}C^{-1}\e^{-(r_1+r_2)e(\alpha)}
\P(\bomega_1\bxi\bomega_2)^{1-\alpha}
\wP(\bomega_1\bxi\bomega_2)^\alpha\\
&\ge\sum_{\bomega_i\in C_i}C^{-1}C_\tau\lambda_0
\left(\e^{-r_1e(\alpha)}\P(\bomega_1)^{1-\alpha}\wP(\bomega_1)^\alpha\right)
\left(\e^{-r_2e(\alpha)}\P(\bomega_2)^{1-\alpha}\wP(\bomega_2)^\alpha\right).
\end{align*}
The upper bound in~\eqref{sick-0} further gives
\begin{align*}
(1+\tau)\sup_{t\ge0}\bar\QQ(\cC_1\cap\phi^{-t}(\cC_2))
&\ge\sum_{t=r_1}^{r_1+\tau}\bar\QQ(\cC_1\cap\phi^{-t}(\cC_2))\\
&\ge\sum_{\bomega_i\in C_i}C^{-3}C_\tau\lambda_0\bar\QQ(\bomega_1)\bar\QQ(\bomega_2)
=C^{-3}C_\tau\lambda_0\bar\QQ(\cC_1)\bar\QQ(\cC_2),
\end{align*}
from which we conclude that the lower bound
$$
\sup_{t\ge0}\bar\QQ(\cC_1\cap\phi^{-t}(\cC_2))
\ge C\,\bar\QQ(\cC_1)\bar\QQ(\cC_2)
$$
holds for any cylinder subsets $\cC_1,\cC_2\subset\Omega$. Since cylinder sets
generate the Borel $\sigma$-field $\cF$, given $\cB_1,\cB_2\in\cF$ and $\epsilon>0$
there are two cylinders sets $\cC_1$ and $\cC_2$ such that 
$\bar\QQ(\cB_i\Delta\cC_i)<\epsilon$. It follows that
$\bar\QQ(\phi^{-t}(\cB_2)\Delta\phi^{-t}(\cC_2))
=\bar\QQ(\phi^{-t}(\cB_2\Delta\cC_2))<\epsilon$ and hence
$$
|\bar\QQ(\cB_1\cap\phi^{-t}(\cB_2))-\bar\QQ(\cC_1\cap\phi^{-t}(\cC_2))|<2\epsilon,
$$
for any $t\ge0$. Accordingly,
$$
\sup_{t\ge0}\bar\QQ(\cB_1\cap\phi^{-t}(\cB_2))
\ge C\,\bar\QQ(\cC_1)\bar\QQ(\cC_2)-2\epsilon
\ge C\,(\bar\QQ(\cB_1)-\epsilon)(\bar\QQ(\cB_2)-\epsilon)-2\epsilon,
$$
and by an appropriate choice of $\epsilon>0$ we can achieve
$$
\sup_{t\ge0}\bar\QQ(\cB_1\cap\phi^{-t}(\cB_2))\ge\frac{C}2\,\bar\QQ(\cB_1)\bar\QQ(\cB_2).
$$
It follows that if  
$\bar\QQ(\cB_1)>0$ and $\bar\QQ(\cB_2)>0$, then there exists $t$ such that 
$\bar\QQ(\cB_1\cap\phi^{-t}(\cB_2))>0$ and Theorem~1.5 in~\cite{Wa} implies that
$\bar\QQ$ is $\phi$-ergodic.\hfill\qed

\noindent{\bf Remark.} Since all limit points $\bar\QQ$ of the sequence 
$\{\QQ_{(T)}\}_{T\ge1}$ satisfy the estimate~\eqref{sick-0}, a standard relative entropy 
argument yields that they are all mutually absolutely continuous. Hence, if one of 
them is ergodic, they all coincide and $\bar\QQ=\lim_{T\to\infty}\QQ_{(T)}$. We will not 
make use of this observation in the sequel.

\medskip
We are now ready to prove Theorem~\ref{th-2}. The case $\alpha\in\{0,1\}$ 
follows from Part~(8) of Theorem~\ref{th-1}. Thus, we need only to consider 
$\alpha\in]0,1[$. We start by noticing that if $\bar\QQ$ is as in Lemma~\ref{sss}, then
the estimates~\eqref{sick-0} imply
\begin{align*}
\varsigma(\bar\QQ^{(\alpha)})&=-\lim_{T\to\infty}\frac1T
\sum_{\bomega\in\Omega_T}
\bar\QQ(\bomega)\log\P(\bomega)^{1-\alpha}\wP(\bomega)^\alpha\\
&\lesseqgtr-\lim_{T\to\infty}\frac1T\sum_{\bomega\in\Omega_T}
\bar\QQ(\bomega)\log\left(C^{\pm1}\e^{Te(\alpha)}\bar\QQ(\omega)\right)\\
&=-e(\alpha)-\lim_{T\to\infty}\frac1T\sum_{\bomega\in\Omega_T}
\bar\QQ(\bomega)\log\bar\QQ(\omega)\\
&=-e(\alpha)+h_\phi(\bar\QQ),
\end{align*}
and hence $\bar\QQ\in\cP_\eq(\alpha)$. To show that $\bar\QQ$ is the only element of
$\cP_\eq(\alpha)$, let $\QQ\in\cP_\eq(\alpha)$. The upper bound in~\eqref{sick-0} yields
\[
S(\QQ_T|\bar\QQ_T)=\sum_{\bomega\in\Omega_T}
\QQ(\bomega)\log\frac{\QQ(\bomega)}{\bar\QQ(\bomega)}
\le-S(\QQ_T)+\log C+Te(\alpha)+T\varsigma(\QQ^{(\alpha)}),
\]
and hence
\beq
0\le\limsup_{T\to\infty}\frac1TS(\QQ_T|\bar\QQ_T)
\le -h_\phi(\QQ)+e(\alpha)+\varsigma(\QQ^{(\alpha)})=e(\alpha)-f(\QQ^{(\alpha)})=0.
\label{yes}
\eeq
Combining Lemmata~\ref{key-simple-1} and~\ref{sss}, we derive
\beq
\bar\QQ_{T+T'}(\bomega\bomega')
\le C^3\lambda_0^{-2}\bar\QQ_T(\bomega)\bar\QQ_{T'}(\bomega'),
\label{perhaps}
\eeq
for all $T,T'\ge1$, $\bomega\in\Omega_T$ and $\bomega'\in\Omega_{T'}$. Writing
$$
a_T=S(\QQ_T|\bar\QQ_T)=-S(\QQ_T)-\QQ[\log\bar\QQ_T],
$$
we deduce from~\eqref{perhaps} and the subadditivity of entropy that
$$
a_{T+T'}\ge a_T+a_{T'}-K,
$$
where $K=\log(C^3\lambda_0^{-2})>0$. Fekete's lemma and~\eqref{yes} then give
$$
0=\lim_{T\to\infty}\frac1T a_T=\sup_{T\ge 1}\frac1T(a_T-K),
$$
from which we conclude that $S(\QQ_T|\bar\QQ_T)\le K$ for all $T\ge1$.
By the lower semicontinuity of the relative entropy we get
$$
S(\QQ|\bar\QQ)\leq\liminf_{T\to\infty}S(\QQ_T|\bar\QQ_T)\le K,
$$
and so $\QQ\ll\bar\QQ$. Since 
$\QQ,\bar\QQ\in\cPi$ and $\bar\QQ$ is $\phi$-ergodic, we have $\QQ=\bar\QQ$. Hence, 
$\cP_\eq(\alpha)$ is a singleton for all $\alpha \in ]0,1[$, and the differentiability 
of $e(\alpha)$ follows from Part~(6) of Theorem~\ref{th-1}. The proof of
Theorem~\ref{th-2} is complete.
%%%%%%%%%%%%%%%%%
\subsection{Proof of Proposition \ref{A-C-1}}
We argue by contradiction. If the statement is not true then, for each $t\ge0$,
there exist $\bomega_t,\bnu_t\in\Omega_{\rm fin}$ such that
$$
\max_{\bxi\in\Omega_{\rm fin}\atop |\bxi|\le t}
\frac{\P(\bomega_t\bxi\bnu_t)\wP(\bomega_t\bxi\bnu_t)}
{\P(\bomega_t)\P(\bnu_t)\wP(\bomega_t)\wP(\bnu_t)}\le \frac1{1+t},
$$
and hence
$$
\lim_{t\to\infty}\frac{\P(\bomega_t\bxi\bnu_t)\wP(\bomega_t\bxi\bnu_t)}
{\P(\bomega_t)\P(\bnu_t)\wP(\bomega_t)\wP(\bnu_t)}=0
$$
for all $\xi\in\Omega_{\rm fin}$.
In terms of the OR process $(\widehat{\cJ},\widehat{\rho}\,)$, this can be rewritten as
\[
\lim_{t\to\infty}
\frac{\tr\left((\Phi_{\bomega_t}^\ast\otimes\widehat{\Phi}_{\bomega_t}^\ast)
[\rho\otimes\widehat{\rho}\,]
(\Phi_{\bxi}\otimes\widehat{\Phi}_{\bxi})\left[
(\Phi_{\bnu_t}\otimes\widehat{\Phi}_{\bnu_t})[\one\otimes\one]\right]\right)}
{\tr\left((\Phi_{\bomega_t}^{\ast}\otimes\widehat{\Phi}_{\bomega_t}^\ast)
[\rho\otimes\widehat{\rho}\,]\right)
\tr\left((\rho\otimes\widehat{\rho}\,)(\Phi_{\bnu_t}\otimes\widehat{\Phi}_{\bnu_t})
[\one\otimes\one]
\right)}=0,
\]
where $\Phi_{\bomega}=\Phi_{\omega_1}\circ\cdots\circ\Phi_{\omega_T}$ for 
$\bomega=(\omega_1,\ldots,\omega_T)$. Since
$$
\tr\left((\rho\otimes\widehat{\rho}\,)(\Phi_{\bnu_t}\otimes\widehat{\Phi}_{\bnu_t})
[\one\otimes\one]\right)
\le\tr\left((\Phi_{\bnu_t}\otimes\widehat{\Phi}_{\bnu_t})
[\one\otimes\one]\right),
$$
we also have
$$
\lim_{t\to\infty}
\frac{\tr\left((\Phi_{\bomega_t}^\ast\otimes\widehat{\Phi}_{\bomega_t}^\ast)
[\rho\otimes\widehat{\rho}\,]
(\Phi_{\bxi}\otimes\widehat{\Phi}_{\bxi})\left[
(\Phi_{\bnu_t}\otimes\widehat{\Phi}_{\bnu_t})[\one\otimes\one]\right]\right)}
{\tr\left((\Phi_{\bomega_t}^{\ast}\otimes\widehat{\Phi}_{\bomega_t}^\ast)
[\rho\otimes\widehat{\rho}\,]\right)
\tr\left((\Phi_{\bnu_t}\otimes\widehat{\Phi}_{\bnu_t})
[\one\otimes\one]
\right)}=0.
$$
By passing to a subsequence $t_n\to\infty$, we can assume that the limits
$$
\varrho=
\lim_{n\to\infty}\frac{(\Phi_{\bomega_{t_n}}^\ast\otimes
\widehat{\Phi}_{\bomega_{t_n}}^\ast)
[\rho\otimes\widehat{\rho}\,]}
{\tr\left((\Phi_{\bomega_{t_n}}^\ast\otimes\widehat{\Phi}_{\bomega_{t_n}}^\ast)
[\rho\otimes\widehat{\rho}\,]\right)},\qquad
\varrho'=\lim_{n\to\infty}\frac{(\Phi_{\bnu_{t_n}}\otimes\widehat{\Phi}_{\bnu_{t_n}})
[\one\otimes\one]}
{\tr\left((\Phi_{\bnu_{t_n}}\otimes\widehat{\Phi}_{\bnu_{t_n}})[\one\otimes\one]\right)}
$$
exist and define density matrices on $\cH\otimes\cH$ such that
\[
\tr(\varrho(\Phi_{\bxi}\otimes\widehat{\Phi}_{\bxi})[\varrho'])=0
\]
for all $\xi\in\Omega_{\rm fin}$.
Hence, the positive map $\Psi=\sum_{a\in\cA}\Phi_a\otimes\widehat{\Phi}_a$ satisfies 
\[
\tr (\varrho\Psi^t[\varrho'])=\sum_{\bxi\in\Omega_t}
\tr(\varrho(\Phi_{\bxi}\otimes\widehat{\Phi}_{\bxi})[\varrho'])=0
\]
for all $t\geq 0$, which contradicts the assumption that $\Psi$ is irreducible; 
see~\cite[Lemma~2.1]{EHK}. \hfill\qed

%%%%%%%%%%%%%%%%%%%%%%%%%%%%%%%%%%%%%%%%%%%%%%
\subsection{Proof of Theorem \ref{th-3}}

Lemma~\ref{key-simple-1} and Assumption~\AssumptionD{} give that there exist  constants $c$ and $c(\alpha)$ 
such that 
for all $T,T'\geq 1$, $\alpha\in\rr$ and $\QQ\in\cPi$ 
$$
\begin{array}{rcccl}
e_T(\alpha)+e_{T'}(\alpha)-c(\alpha)&\leq& e_{T+T'}(\alpha)&\leq
&e_T(\alpha)+e_{T'}(\alpha)+c(\alpha),\\[8pt]
\QQ[-\log\P_T]+\QQ[-\log\P_{T'}]-c&\le&\QQ[-\log\P_{T+T'}]&\le&
\QQ[-\log\P_T]+\QQ[-\log\P_{T'}]+c.
\end{array}
$$
By Fekete's lemma the limits $e(\alpha)=\lim_{T\to \infty}\frac1T e_T(\alpha)$ 
and $\varsigma(\QQ)=\lim_{T\to \infty}\frac1T\QQ[-\log\P_T]$ exist
and are finite. In particular, the relations
$$
\inf_{T\ge1}\frac1T\left(\QQ[-\log\PP_T]-c\right)=\varsigma(\QQ)
=\sup_{T\ge1}\frac1T\left(\QQ[-\log\PP_T]+c\right),
$$
imply that the map $\cPi\ni\QQ\mapsto\varsigma(\QQ)$ is continuous. Going back to the
proof of Theorem~\ref{th-1}~(5), we can apply Lemma~2.3 in~\cite{CFH} to the 
sequence $f_T=1/(D_0\P^\#_T)$ to complement~\eqref{san2} with the estimate
$$
\liminf_{k\to\infty}Q_{T_k}\left[\frac1{T_k}\log\P^\#_{T_k}\right]
\ge\lim_{T\to\infty}\QQ\left[\frac1T\log\P^\#_T\right],
$$
which, together with~\eqref{san2}, implies
$$
\lim_{k\to\infty}\,(1-\alpha)Q_{T_k}\left[\frac1{T_k}\log\P_{T_k}\right]
+\alpha Q_{T_k}\left[\frac1{T_k}\log\wP_{T_k}\right]=-\varsigma(\QQ^{(\alpha)}).
$$
With this addition, the proof of Parts~(5--6) of Theorem~\ref{th-1} now extend to all 
$\alpha\in\rr$. Finally, to prove that $\cP_\eq(\alpha)$ is a singleton for all 
$\alpha \in \rr$, one follows the arguments of Section~\ref{sec-theend}. Replacing 
Assumption~\AssumptionC{}  with Assumption~\AssumptionD{} one easily shows that 
Lemmata~\ref{lemma-long} and~\ref{sss} also extend to all $\alpha\in\rr$.
The details are considerably simpler than in Section~\ref{sec-theend} and we leave them 
to the interested reader. 

%%%%%%%%
\subsection{Proof of Proposition \ref{never-1}}
  We argue by contradiction: if the statement
is not true, we can find  two sequences  
$\bomega_n$ and $\bnu_n$ in $\Omega_{\rm fin}$ such that 
\beq
\lim_{n\to\infty}\frac{\P(\bomega_n\bnu_n)}{\P(\bomega_n)\P(\bnu_n)}=0.
\label{theend}
\eeq
Passing to subsequences one easily deduces from~\eqref{theend} that there exist two 
density matrices $\varrho$ and $\varrho'$ on $\cH$ and some $a\in\cA$ such that
$\bnu_n=a\tilde{\bnu}_n$ and
$$
\varrho=\lim_{n\to\infty}
\frac{\Phi^\ast_{\bomega_n}(\rho)}{\tr(\Phi^\ast_{\bomega_n}(\rho))},
\qquad
\varrho'=\lim_{n\to\infty}
\frac{\Phi_{\tilde{\bnu}_n}(\one)}{\tr(\Phi_{\tilde{\bnu}_n}(\one))}.
$$ 
It follows that $\tr(\varrho\Phi_a(\varrho'))=0$ which contradicts the assumption that 
$\Phi_a$ is positivity improving.\hfill\qed
%%%%%%%%%%%%%%%%%%%%%%%%%%%%%%%%%%%%%%%%%%%%%%
\subsection{Proof of Theorem \ref{thm-chernoff}}

The observation 
\[
c_T=\frac12\sum_{\omega\in\Omega_T}\min\left(\P_T(\omega),\wP_T(\omega)\right)\leq
\frac12\sum_{\omega\in\Omega_T}\P_T(\omega)^{1/2}\wP_T(\omega)^{1/2},
\]
yields $\log c_T\le e_T(\tfrac12)-\log2$, and Theorem~\ref{th-1}~(1) gives Part~(1). 
To prove Part~(2), note that Eq.~\eqref{NPeq} implies
\[
c_T\geq\frac12\P_T(\ubar \cT_T^c)
=\frac12\P_T\left(\left\{\omega\in\Omega_T\,|\, \sigma_T(\omega)<0\right\}\right).
\]
Since the function $e(\alpha)$ has a global minimum at $\alpha=\tfrac12$, one has 
$0\in\partial e(\tfrac12)$, and Theorem~\ref{ldp-1}~(2) gives 
\[
\liminf_{T\to\infty}\frac1T\log c_T
\geq -I\left(-\partial^+e\left(\tfrac12\right)\right).
\]
Part~(2) thus follows from Eq.~\eqref{iloc1}. Finally, under Assumption~\AssumptionC,
Theorem~\ref{th-2}~(2) and the above argument yield $e'(\tfrac12)=0$, and so 
Part~(3) follows from Parts~(1) and~(2).

%%%%%%%%%%%%%%%%%%%

\end{document}